\pdfoutput=1
\documentclass[11pt]{article}

\usepackage[]{ACL2023}
\usepackage{times}
\usepackage{latexsym}

\usepackage[T1]{fontenc}

\usepackage[utf8]{inputenc}

\usepackage{microtype}

\usepackage{inconsolata}

\usepackage{graphicx}
\usepackage{subfigure}
\usepackage{amssymb}
\usepackage{amsmath}
\usepackage{multirow}
\usepackage{tabularx}
\usepackage{booktabs,amsfonts,dcolumn}
\usepackage{bigstrut}

\title{CLAPSpeech: Learning Prosody from Text Context with Contrastive Language-Audio Pre-training}

\author{ 
	Zhenhui Ye\thanks{~ Equal contribution.} \\ zhenhuiye@zju.edu.cn \\ Zhejiang University 
	\And Rongjie Huang\footnotemark[1] \\ rongjiehuang@zju.edu.cn \\Zhejiang University 
	\And Yi Ren \\ ren.yi@bytedance.com \\ Bytedance  
	\AND
	 Ziyue Jiang \\ jiangziyue@zju.edu.cn \\ Zhejiang University \\ 
	\And Jinglin Liu \\ liu.jinglin@bytedance.com \\ ByteDance \\ 
	\And Jinzheng He \\ jinzhenghe@zju.edu.cn \\ Zhejiang University \\ 
	\AND Xiang Yin \\ yixiang.stephen@bytedance.com \\ Bytedance 
	\And Zhou Zhao\thanks{ ~  Corresponding author.} \\ zhaozhou@zju.edu.cn\\ Zhejiang University \\ 
}

\begin{document}
	\maketitle
	\begin{abstract}
		Improving text representation has attracted much attention to achieve expressive text-to-speech (TTS). However, existing works only implicitly learn the prosody with masked token reconstruction tasks, which leads to low training efficiency and difficulty in prosody modeling. We propose CLAPSpeech, a cross-modal contrastive pre-training framework that explicitly learns the prosody variance of the same text token under different contexts. Specifically, 1) We encourage the model to connect the text context with its corresponding prosody pattern in the joint multi-modal space with the elaborate design of the encoder inputs and contrastive loss; 2) We introduce a multi-scale pre-training pipeline to capture prosody patterns in multiple levels. We show how to incorporate CLAPSpeech into existing TTS models for better prosody. Experiments on three datasets not only show that CLAPSpeech could improve the prosody prediction for existing TTS methods, but also demonstrate its generalization ability to adapt to multiple languages and multi-speaker TTS.	We also deeply analyze the principle behind the performance of CLAPSpeech. Ablation studies demonstrate the necessity of each component in our method. Source code and audio samples are available at \url{https://clapspeech.github.io}.
		
	\end{abstract}
	
	\section{Introduction}
	
	\begin{figure*}[!t]
		\centering
		\includegraphics[width=1.0\linewidth]{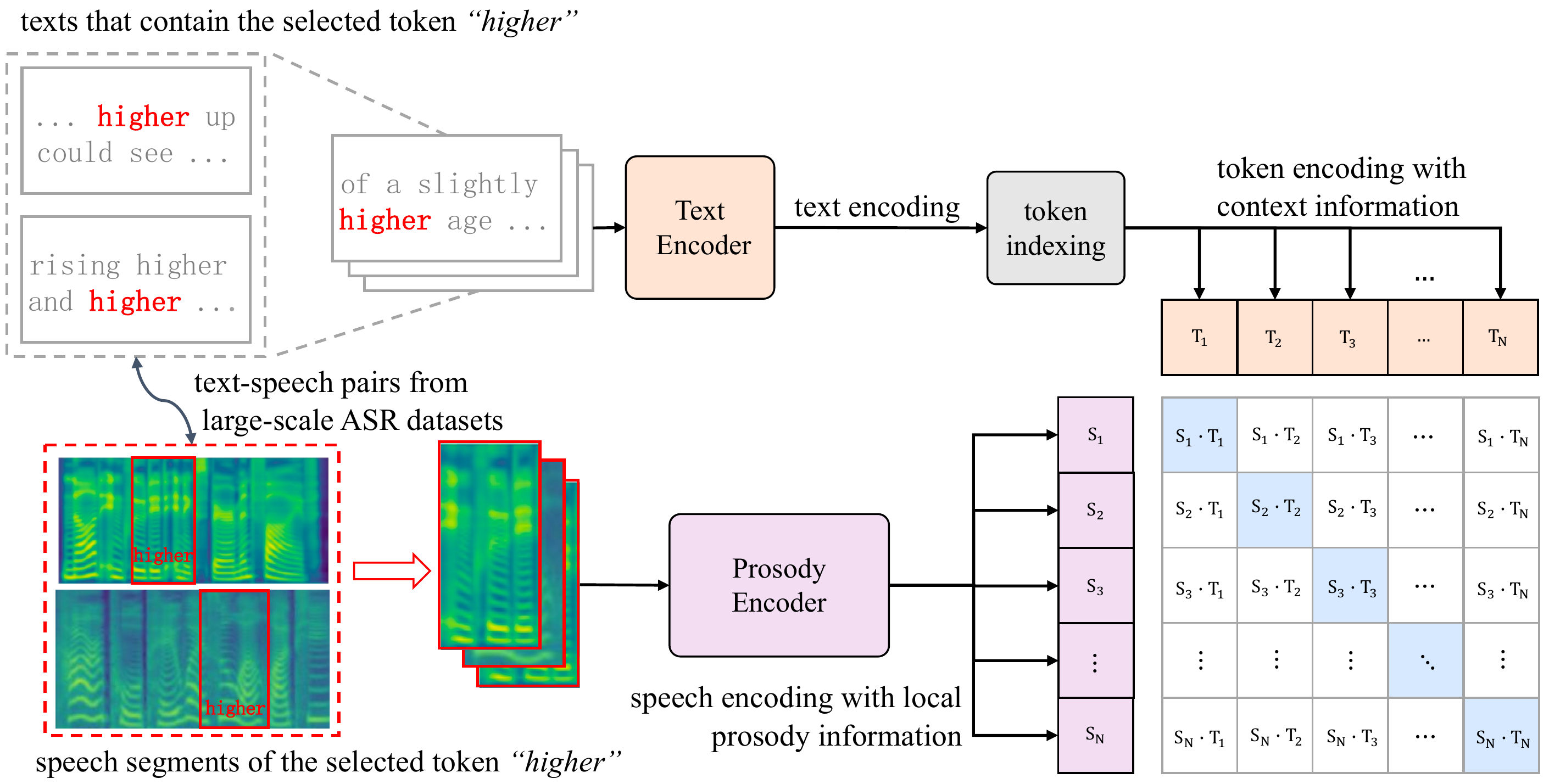}
		\caption{The contrastive pre-training process of CLAPSpeech. For clarity, we only show the word-level pre-training here. Note that we also perform a phoneme-level pre-training.}
		\vspace{-3mm}
		\label{fig:train_clip}
	\end{figure*}
	
	With the development of deep learning, the audio quality of modern TTS systems has been improved, yet prosody modeling is still a challenging problem. Previous works on expressive TTS have utilized external variation predictors (prediction-based, PB) \cite{ren2021fastspeech2} and variational generative models (variation-based, VB) \cite{kim2020glow,liu2022diffsinger} to inject prosody variance into the TTS model. Another popular direction is to learn better text representation for prosody prediction \cite{tan2021survey}. However, the existing text representation learning methods for TTS are either based on the masked language model task \cite{devlin2019bert,jia2021png,chen2021speech} (i.e., learn a BERT-like large language model on a text corpus) or masked acoustic model task \cite{chen2020mam,bai2022a3t} (i.e., reconstruct the masked mel-spectrogram based on the input text), which result in two disadvantages. Firstly, they only implicitly learn prosody with reconstruction losses, which distracts the model from improving the prosody modeling. Secondly, they do not decouple the pronunciation space and prosody space, which leads to low training efficiency and a waste of model capacity. We perform a case study in Section \ref{sec:discuss_self_similarity}, in which we can see that previous text representation used in TTS cannot capture the prosody variance under different text contexts.
	
	Technically, prosody can be regarded as the pitch and duration variance of the same token under different conditions (such as text contexts and speakers) \cite{tan2021survey}. This paper mainly studies the prosody correlated to the text context. For instance, for the same word \textit{"higher"}, saying \textit{"higher up"} or \textit{"slightly higher"} can lead to different prosodies. Inspired by recent cross-modal contrastive learning works in the text-to-image task \cite{radford2021clip,elizalde2022clap}, we propose a contrastive learning method that connects the text context and the high-level prosody pattern in the text-speech joint multi-modal space, namely \textbf{C}ontrastive \textbf{L}anguage-\textbf{A}udio \textbf{P}re-Training for Text-to-\textbf{Speech} (CLAPSpeech). Specifically, we learn a text encoder to predict the prosody from the text context and a prosody encoder to extract the ground-truth (GT) prosody from the speech segment of the selected token. During training, we select $N$ text-speech pairs that contain the same pronounceable token (e.g., the word \textit{"higher"} or phoneme \textit{"AE0"}). By aligning the text token with its corresponding prosody (extracted from GT speech) and pushing away the prosody representation from other text contexts, the text encoder is encouraged to extract prosody from the text context. An intuitive example of pre-training CLAPSpeech can be found in Figure \ref{fig:train_clip}. We also observe that the prosody pattern can be expressed at multiple levels. Therefore, we propose a multi-scale pre-training framework that learns two CLAPSpeech models to capture the prosody information at the phoneme and word levels, respectively. After the pre-training stage, our CLAPSpeech can be regarded as a plug-in text encoder applicable to all TTS models to provide fine-grained prosody representation. 
	
	To prove the effectiveness and generalizability of our approach, we use two large-scale automatic speech recognition (ASR) datasets (LibriSpeech \cite{panayotov2015librispeech} for English and WenetSpeech \cite{zhang2022wenetspeech} for Chinese) to pre-train the CLAPSpeech model. The pre-trained text encoder of CLAPSpeech is then plugged into prediction/variation-based TTS baselines to demonstrate the improvement of CLAPSpeech to the existing expressive TTS systems. We then evaluate the performance on three TTS datasets, including one single-speaker English dataset, one single-speaker Chinese corpus, and one multi-speaker English dataset. Experiments on all datasets show that CLAPSpeech improves the prosody of the TTS models and outperforms previous representation learning methods. 
	
	To summarize, CLAPSpeech has three prominent advantages: 1) It can provide better prosody representation than previous representation learning methods with a much smaller model scale, thanks to its contrastive objective that explicitly learns the prosody. 2) The text representation of CLAPSpeech can be conveniently used in existing TTS systems, only with a minor modification of the front-end network architecture. 3) We also show its potential applications such as fine-grained prosody transfer in Section \ref{sec:prosody_transfer}.

	
	\section{Related Work}
	\subsection{Expressive TTS}
	In the past few years, modern neural TTS has made significant progress in high practicality and audio quality \cite{ren2019fastspeech,kim2020glow,elias2021parallel,miao2021efficienttts,kim2021conditional,donahue2020end,jiang2022dict}. However, modeling expressive prosody given the plain input text is still challenging. To achieve expressive TTS, one common practice is to use a reference encoder and style tokens \cite{wang2018style,jia2018transfer}. But it is difficult to select appropriate reference audios during inference \cite{tan2021survey}. Other works seek to improve prosody modeling with advanced network designs, which can be categorized into two classes: (1) the prediction-based (PB) TTS systems \cite{ren2021fastspeech2} learn several external predictors to predict the prosody attributes such as pitch contour, duration, and energy; (2) the variation-based (VB) TTS systems leverage variational auto-encoder (VAE) \cite{ren2021portaspeech} or normalizing flow \cite{kim2020glow} to model the prosody in the latent space. 
	

	
	There are also some works that explore providing better text presentation with rich prior knowledge to help the prosody prediction. For instance, \citet{liu2021graphspeech} and \citet{ye2022syntaspeech} incorporate syntax information through dedicated modeling methods such as graph networks. Representation learning methods for text pre-training and speech pre-training also show improvements in the prosody of TTS. We will discuss the representation learning works for TTS in the next section.
	
	\subsection{Representation Learning for TTS}
	Self-supervised pre-training methods have been leveraged in TTS to enhance text processing or speech generation capabilities \cite{chung2019semi,zhang2019joint}. Some early works \cite{wang2015word} use pre-trained word embeddings to improve the robustness of TTS systems. Recently, some works explore incorporating pre-trained large masked language models (MLMs) \cite{devlin2019bert, chen2021speech, jia2021png} to enjoy the rich semantic information learned from the web-scale text corpus. However, the above-mentioned works only focus on the text space, it is challenging for them to model expressive prosody considering the models are unaware of the high variable prosody patterns in the speech space. There are several inspiring speech representation learning methods in ASR. \citet{baevski2020wav2vec} and \citet{hsu2021hubert} utilize masked continuous speech features to predict predetermined cluster assignments. As for TTS, ProsoSpeech \cite{ren2022prosospeech} designs a word-level vector quantization bottleneck to extract discrete prosody representation from speech. Masked acoustic model (MAM) \cite{chen2020mam} proposes to learn a speech encoder that generates continuous speech (prosody) representations. Specifically, during training they replace a span of speech spectrogram with mask tokens and learn to recover the masked spectrogram without text conditions. A$^3$T \cite{bai2022a3t} additionally learns a text encoder as auxiliary information for MAM to reconstruct the masked mel-spectrogram. 
	
	The difference between CLAPSpeech and previous representation works in TTS is obvious: While previous works implicitly learn the prosody information with the masked token reconstruction task, CLAPSpeech is the first work that utilizes the cross-modal contrastive learning to explicitly learn the context-correlated prosody, which leads to better prosody prediction and more efficient usage of model capacity.
	
	
	\section{CLAPSpeech}
	We propose CLAPSpeech, a cross-modal contrastive learning approach to provide better text representation for prosody prediction in TTS. As shown in Figure \ref{fig:train_clip}, CLAPSpeech comprises a text encoder and a prosody encoder, whose training objective is to connect the text token and the speech segment in the joint prosody space. In this section, we first design the network structure and input features of these two encoders. These elaborate designs enable the text encoder to effectively process the text context and ensure that the prosody encoder focuses on extracting the high-level prosody pattern from the speech segment while eliminating other variables, such as timbre. Then we introduce the multi-scale contrastive pre-training framework, which enables CLAPSpeech to capture prosody in both phoneme and word levels. Finally, we show how the pre-trained text encoder of CLAPSpeech can be conveniently plugged into modern TTS systems to improve prosody prediction. We describe these designs in detail in the following subsections and provide more technical details in Appendix \ref{appendix:details_of_models}.
	
	\subsection{Text Encoder and Prosody Encoder}
	
	\begin{figure}
		\centering
		\subfigure[text encoder]{
			\begin{minipage}[t]{0.5\linewidth}
				\centering
				\includegraphics[width=0.99\linewidth]{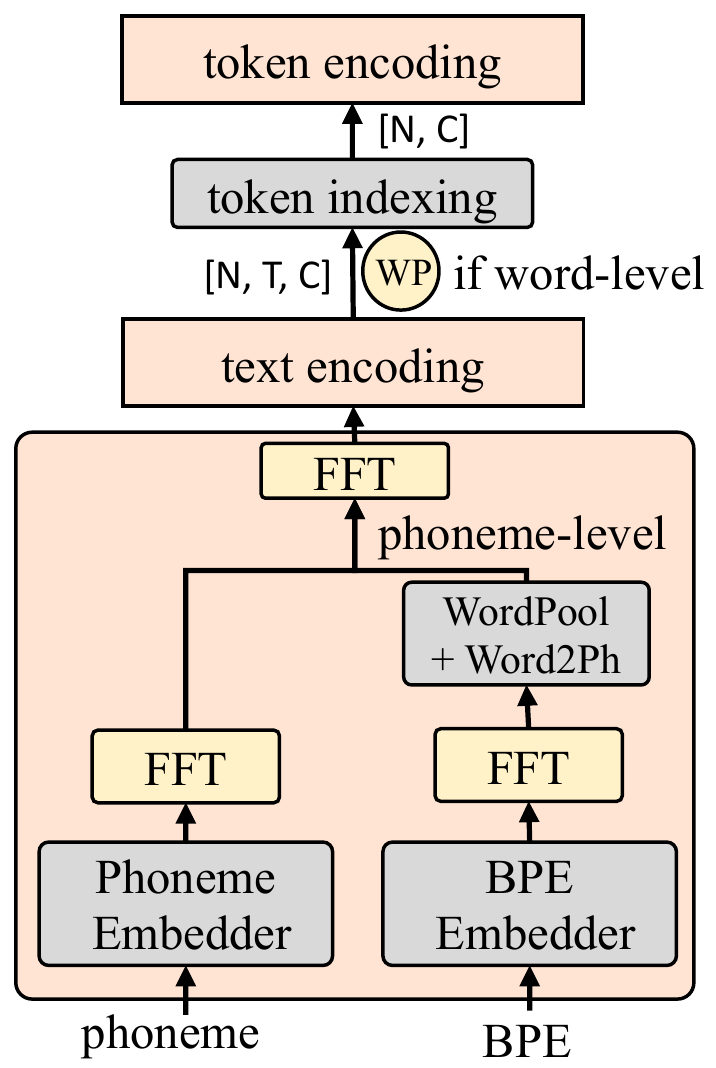}
			\end{minipage}
			\label{fig:text_encoder}
		}%
		\subfigure[prosody encoder]{
			\begin{minipage}[t]{0.5\linewidth}
				\centering
				\includegraphics[width=0.99\linewidth]{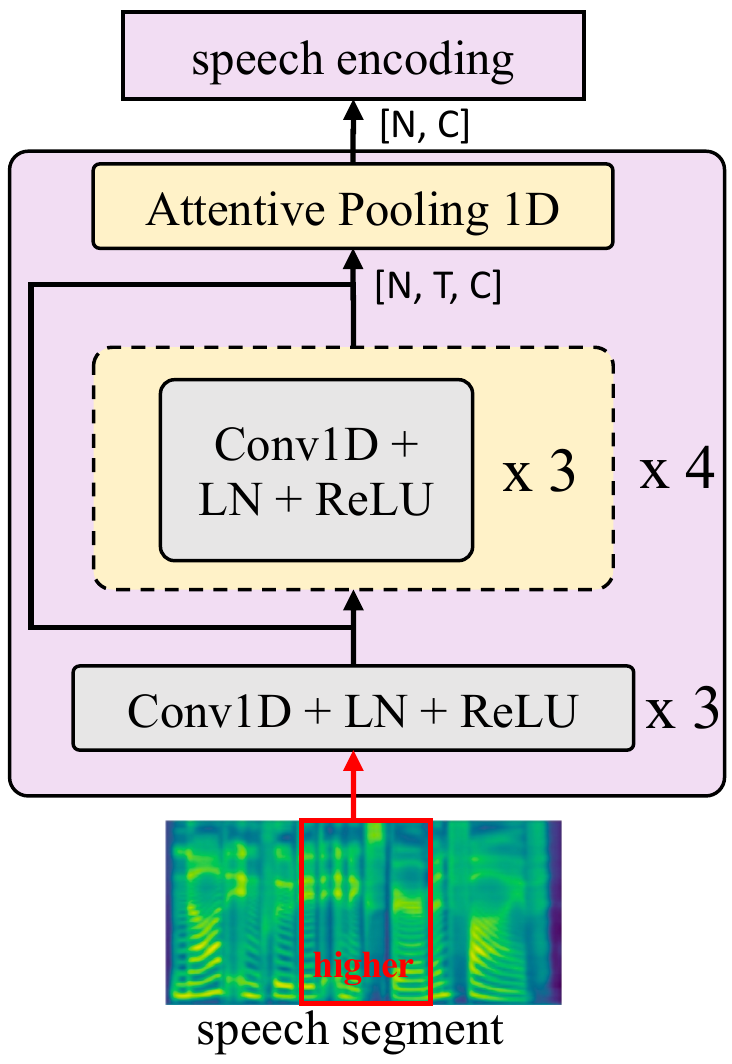}
			\end{minipage}
			\label{fig:prosody_encoder}
		}
		\caption{The text / prosody encoder of CLAPSpeech. In subfigure (a), "WP" and "Word2Ph" denotes word pooling and Word2Ph expanding operation, which are illustrated in Figure \ref{fig:wp_and_word2ph}.}
		\label{fig:text_encoder_and_prosody_encoder}
	\end{figure}
	
	The prosody of the same pronounceable token\footnote{such as the phoneme \textit{"AE0"} or the word \textit{"higher"}.} varies in different text contexts. CLAPSpeech aims to model the correlation between the text context and the high-level prosody pattern. To this end, we design a text encoder and a prosody encoder to construct a text-speech multi-modal prosody embedding space.
	
\begin{figure}
		\centering
		\subfigure[Word Pooling]{
			\begin{minipage}[t]{0.49\linewidth}
				\centering
				\includegraphics[width=0.99\linewidth]{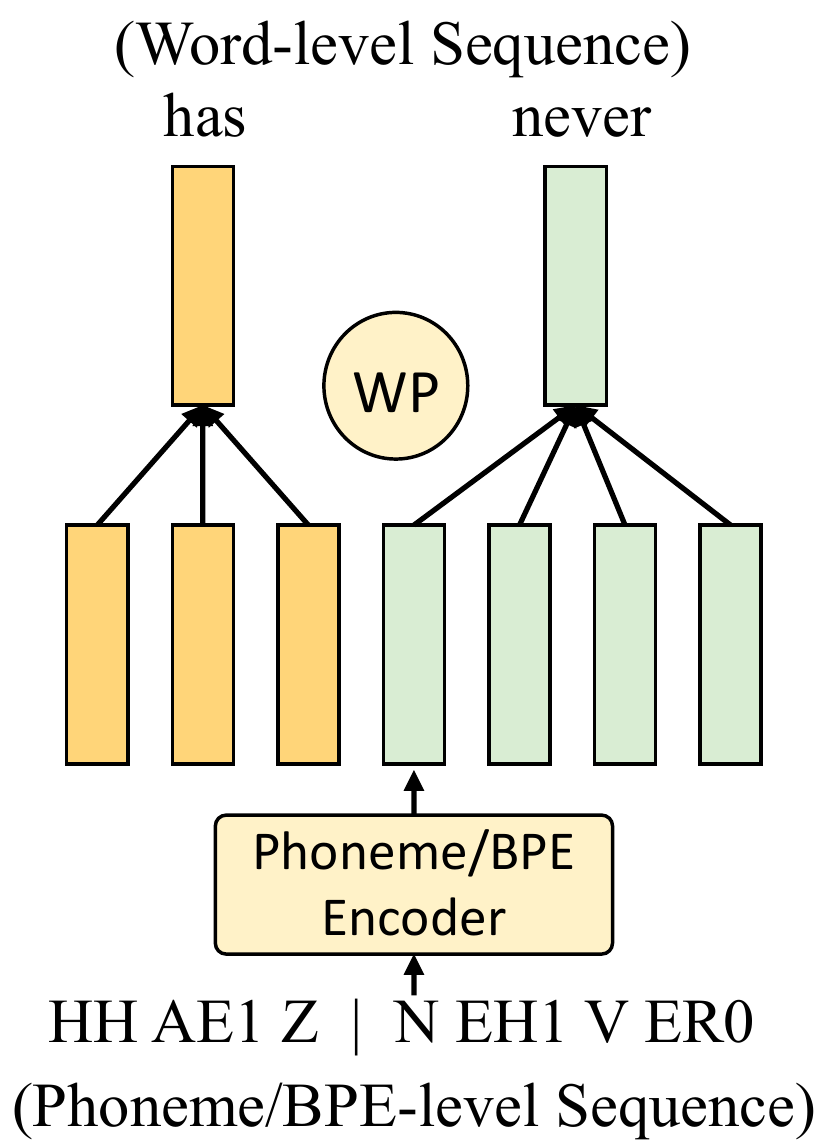}
			\end{minipage}
			\label{figure:word_pool}
		}%
		\subfigure[Word2Ph Expanding]{
			\begin{minipage}[t]{0.46\linewidth}
				\centering
				\includegraphics[width=0.99\linewidth]{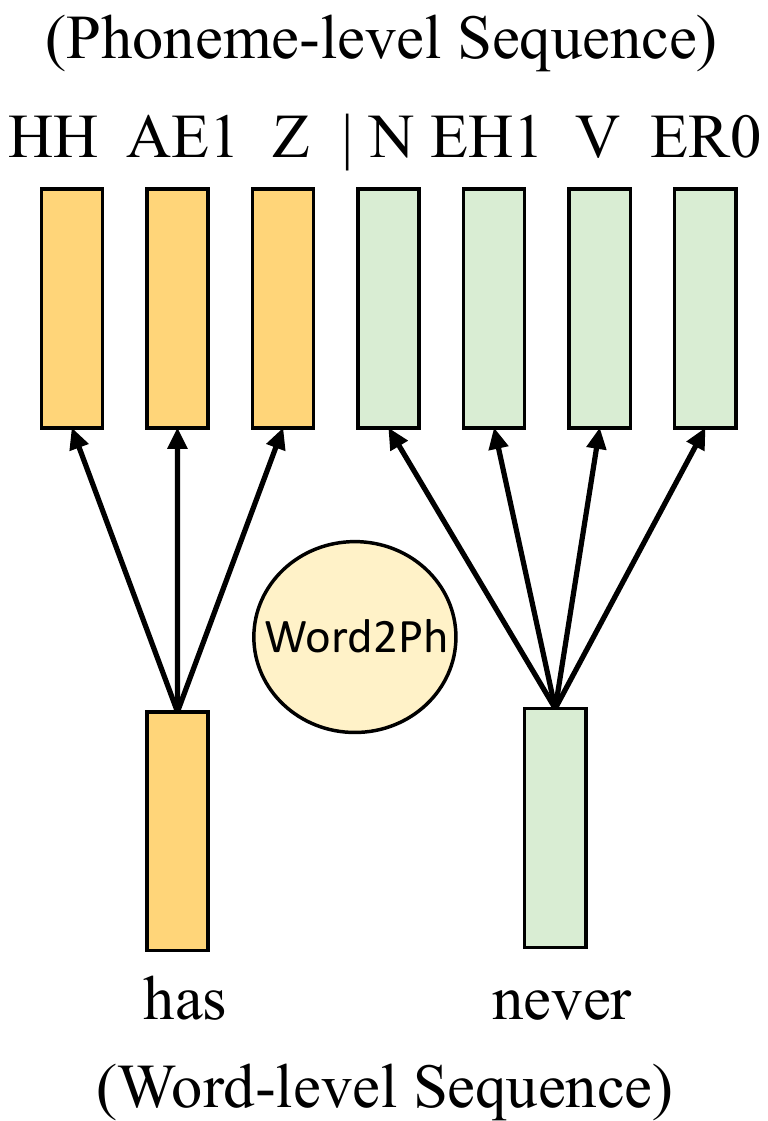}
			\end{minipage}
			\label{figure:word2ph}
		}
		\caption{The word pooling and word2ph expanding operation.}
		\label{fig:wp_and_word2ph}
	\end{figure}

	As shown in Figure \ref{fig:text_encoder}, the text encoder uses phoneme and byte pair encoding (BPE) \cite{shibata1999byte} of the input text as the input. The phoneme and BPE sequence help the model extract the prosody pattern related to phonological habits (such as the linking phenomenon in English) and semantic information (which may imply different emotional overtones), respectively. The network structure of the text encoder is composed of several Feed Forward Transformers (FFT) \cite{vaswani2017attention}, which have proven the robustness in processing long text sequences in TTS models. Specifically, we learn two independent FFT blocks to process the phoneme and BPE sequences, respectively. This way, the phoneme FFT block could model the phonological habits in phonetic space, and the BPE FFT block could extract the semantic information. One difficulty is fusing the phoneme and BPE sequence of mismatched length. Instead of concatenating these two sequences in the time axis, we use word-level pooling (WP) from \citet{ren2021portaspeech} to process the BPE encoding to the word level, then expand it to the phoneme level (namely the \textit{word2ph} operation). To be specific, as shown in Figure \ref{figure:word_pool}, the WP operation averages the phoneme hidden states inside each word according to the word boundary, and the word2ph operation repeats the word hidden states for each phoneme insides the word boundary as illustrated in Figure \ref{figure:word2ph}.
	
	Once the phoneme sequence and BPE seqneuce is fused, we then use an additional FFT block to fuse the aligned phoneme  and BPE encoding to get the final phoneme-level text encoding. During the pre-training phase, since only one selected token is analyzed, we index from the phoneme-level text encoding to obtain the encoding of the selected token (namely the \textit{token encoding} in Figure \ref{fig:text_encoder}) and then linearly project it into the multi-modal embedding space. During the TTS phase, the phoneme-level output of the text encoder can be conveniently utilized as auxiliary features for TTS systems, which we will discuss in Section \ref{sec:clapspeech_in_tts}.
	
	The prosody encoder aims to extract prosody patterns from the GT speech segment of the selected token. Therefore, we clip the mel-spectrogram with the word boundary\footnote{ We extract word boundary with a forced alignment tool.} as the input speech feature. Then the prosody encoder processes the input mel-spectrogram into a global encoding to be connected with the token encoding. Note that the clipped speech segment only contains the local prosody information for the selected token without leaking any contextual information. Thanks to the contrastive learning setting, the extracted global prosody encoding is disentangled from phonetic and speaker space: 1) since the positive sample and negative samples belong to the same pronounceable token, the phonetic information is eliminated; 2) as the speaker information is not provided to the text encoder\footnote{We assume that text and speaker are independent of each other (no correlation between them) in our dataset.}, the prosody encoder will filter out speaker information to maximize the prosody information in the output features during training.
	This way, by connecting the context-aware text encoding with the context-unaware mel encoding, on the one hand, the prosody encoder learns to extract the high-level prosody information from the speech segment; on the other hand, the text encoder is encouraged to utilize the text context to predict the prosody extracted by the prosody encoder. As shown in Figure \ref{fig:prosody_encoder}, we use ResNet-50 \cite{he2016deep} as the backbone of the prosody encoder due to its robustness. We make several modifications to the original version: 1) to better process the mel-spectrogram, we use 1D convolution with layer normalization to build the fundamental residual block; 2) to handle the speech segment of dynamic lengths, we use an attentive pooling layer from \citet{radford2021clip} to aggregate the output feature map of the ResNet. 

	\subsection{Multi-scale Contrastive Pre-training }
	\label{sec:multi-scale-pretraining}
	The key idea of CLAPSpeech is to model the prosody variance of the same text token under different contexts. Therefore, to construct a mini-batch for contrastive pre-training, we randomly select a text token, then sample a batch of $N$ text-speech pairs that contain the selected token (one intuitive sample is shown in Figure \ref{fig:train_clip}, where we sample the text-speech pairs that contain the word \textit{"higher"}). To better extract prosody variance at the phoneme and word level, we introduce a multi-scale contrastive training framework. To be specific, we learn two CLAPSpeech models for phoneme-level and word-level text tokens, respectively.
	
	For clarity, we first illustrate the training process of phoneme-level CLAPSpeech. Let the text context that contains the selected phoneme token (e.g., \textit{"AE0"}) be represented by $X_{text}$. Let the processed speech segment of the phoneme token be $X_{speech}$ s.t. $X_{speech}\in \mathbb{R}^{F\times T}$, where $F$ is the number of Mel bins and $T$ is the number of time bins. For simplicity, we use $X_{text}$ and $X_{speech}$ to represent a batch of $N$ text-speech pairs.
	
	The text and speech are passed through the text encoder $f_{text}(\cdot)$ and prosody encoder $f_{speech}(\cdot)$, respectively. As can be seen in Figure \ref{fig:text_encoder}, the output of the text encoder $f_{text}(X_{text})$ is the phoneme-level encoding of the input text, hence we index from it to obtain the encoding of the phoneme token $f_{text}(X_{text})_{i_{ph}}$, where $i_{ph}$ denotes the index of the phoneme token in the phoneme-level text sequence. As can be seen in Figure \ref{fig:prosody_encoder}, the output speech encoding $f_{speech}(X_{speech})$ is a global representation of the input speech segment. The output representations are normalized and then linearly projected into the multi-modal embedding space:
	\begin{equation}
		\begin{split}
			T_{ph} = L_{text}(LN(f_{text}(X_{text})_{i_{ph}})) \\
			S = L_{speech}(LN(f_{speech}(X_{speech}))),
		\end{split}
	\end{equation}
	where $T_{ph}\in \mathbb{R}^{N\times C}$ is the phoneme token representation and $S\in \mathbb{R}^{N\times C}$ is the speech representation of channel size $C$. $LN$ means layer normalization, $L_{text}$ and $L_{speech}$ are linear projections.
	
	Now that the text and speech embeddings are comparable, CLAPSpeech is trained to predict which of the $N\times N$ possible text-speech pairings across a batch actually occurred. Specifically, the text encoder and prosody encoder are encouraged to maximize the cosine similarity of the text and speech encoding of the $N$ real pairs in the batch while minimizing the cosine similarity of the embeddings of the $N^2-N$ incorrect pairings. Following \citet{radford2021clip}, we optimize a symmetric cross-entropy loss over these similarity scores:
	\begin{equation}
		\label{eq:clip_ph}
		\mathcal{L}_{ph}=0.5\times({l}_{text}(\tau\cdot C_{ph})+{l}_{speech}(\tau\cdot C_{ph}))
	\end{equation}
	where $C_{ph}\in\mathbb{R}^{N\times N}$ is the cosine similarity matrix between the phoneme token encoding $T_{ph}$ and the speech encoding $S$, measured by $C_{ph}= T_{ph}\cdot S^T$; $\tau$ is a learnable temperature parameter to scale the range of logits; and $l_{k}=\frac{1}{N} \Sigma_{i=0}^{N}\log \textrm{diag}(\textrm{softmax}(C))$ is the cross entropy function along the text and speech axis in $C$. 
	
	The word-level CLAPSpeech can be trained similarly. As shown in Figure \ref{fig:text_encoder}, for the word-level CLAPSpeech, we use word pooling to process the phoneme-level text encoding into word level, then index from it to obtain the word token encoding $T_{word}$. Similar to Equation \ref{eq:clip_ph}, the training loss for word-level CLAPSpeech is formulated as:
	\begin{equation}
		\mathcal{L}_{word}=0.5\times({l}_{text}(\tau\cdot C_{word})+{l}_{speech}(\tau\cdot C_{word}))
	\end{equation}
	where $C_{word}$ is the cosine similarity matrix between the word token encoding $T_{word}$ and the speech encoding $S$. 
	
	\subsection{CLAPSpeech Plugged in TTS Systems}
	\label{sec:clapspeech_in_tts}
\begin{figure}[!t]
		\centering
		\includegraphics[width=0.8\linewidth]{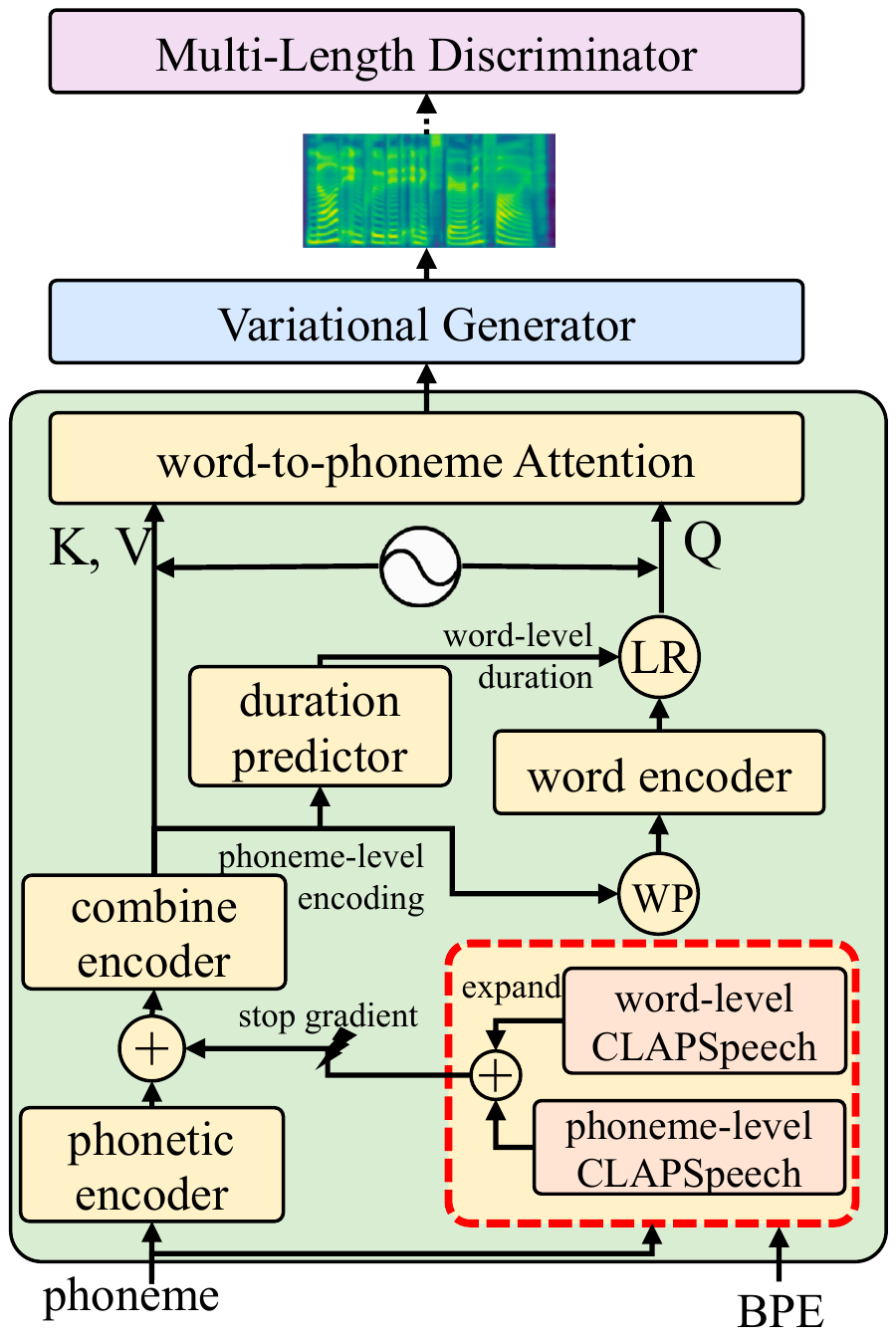}
		\caption{PortaSpeech with CLAPSpeech plugged in.}
		\vspace{-3mm}
		\label{fig:clapspeech_ps_adv}
	\end{figure}

	The text encoder of CLAPSpeech could provide text representation with rich prosody information for the TTS task. Since the generated text representation is at the phoneme level, which is in line with the majority of current TTS models that also utilize phoneme sequence as the text input, CLAPSpeech can be a convenient plugin unit for TTS systems to improve prosody prediction. Specifically, we take a state-of-the-art variation-based TTS system, \textit{PortaSpeech}, as an example. As shown in Figure \ref{fig:clapspeech_ps_adv}, the pre-trained text encoders of CLAPSpeech (marked with a red dashed rectangle) perform as an auxiliary encoder to the original phonetic encoder of PortaSpeech. The phoneme-level outputs of the phonetic encoder and CLAPSpeech text encoder are fused and processed by the following encoder. Note that we fix the parameters of CLAPSpeech text encoders during the training of the TTS system to avoid overfitting. CLAPSpeech can be easily plugged into other TTS systems in a similar way. To demonstrate the universality, we illustrate how to combine CLAPSpeech with a widely-used prediction-based TTS system, \textit{FastSpeech 2}, in Appendix \ref{appendix:clapspeech_fs2}. We additionally adopt multi-length adversarial training in TTS models to improve audio quality. More details about the the adversarial training can be found in Appendix \ref{appendix:multi-length}.
	
	\section{Experiments}
	
	\subsection{Experimental Setup}
	\label{sec:experimental_setup}
	\paragraph{Datasets and Baselines}
	We pre-train CLAPSpeech on two ASR datasets: 1) LibriSpeech \cite{panayotov2015librispeech}, an English database that contains 982 hours of speech from 2484 speakers; 2) WenetSpeech \cite{zhang2022wenetspeech}, a Chinese speech corpus consisting of 10,000 hours of speech\footnote{We filter samples with a correctness confidence level above 0.95, finally get a subset of 1000 hours.}. Then we evaluate the pre-trained CLAPSpeech on three TTS datasets: 1) LJSpeech \cite{ljspeech17}, a single-speaker database that contains 13,100 English audio clips with a total of nearly 24 hours of speech; 2) Biaobei\footnote{\url{https://www.data-baker.com/open source.html}}, a Chinese speech corpus consisting of 10,000 sentences (about 12 hours) from a Chinese speaker; 3) LibriTTS \cite{zen2019libritts}, an English dataset with 149,736 audio clips (about 245 hours) from 1,151 speakers (We only use \textit{train clean360} and \textit{train clean100}). The raw text is transformed into phoneme and BPE sequences using open-sourced tools. The GT mel-spectrograms are generated from the raw waveform with a frame size of 1024 and the hop size of 256. We compare CLAPSpeech against two pre-training baselines (BERT \cite{devlin2019bert} and A$^3$T \cite{bai2022a3t}) in a prediction-based (PB) TTS model, FastSpeech 2, and a variation-based (VB) TTS model, PortaSpeech.
	
	\paragraph{Model Configuration}
	CLAPSpeech consists of a text encoder and a prosody encoder, whose structures are shown in Figure \ref{fig:text_encoder_and_prosody_encoder} and discussed in Section \ref{sec:multi-scale-pretraining}. As for the PB and VB TTS models, we use the same structure in the original papers with an additional multi-length discriminator to improve audio quality. The multi-length discriminator consists of multiple stacked convolutional layers with batch normalization and treats the input spectrogram as images. We put more detailed model configurations in Appendix \ref{appendix:hyper_params}.
	
	\paragraph{Training and Evaluation}
	Our approach is implemented with Pytorch. We pre-train CLAPSpeech on 4 Nvidia 3090Ti GPUs with a batch size of 1,024 text-speech pairs (256 pairs per GPU). We use the Adam optimizer with an initial learning rate of 0.0005. We train the CLAPSpeech model for 640,000 iterations (which takes about 1 week) and follow the cosine learning rate schedule in CLIP. Then we train the TTS models on 1 Nvidia 2080Ti GPU with a batch size of 64 sentences, following the learning rate schedule in \citet{vaswani2017attention}. We use HiFi-GAN \cite{kong2020hifi} as the vocoder. We conduct the mean opinion score (MOS) and comparative mean opinion score (CMOS) evaluation to measure the prosody and audio quality. Details about the subjective evaluation can be found in Appendix \ref{appendix:subjective_evaluation}. As for the objective evaluation, following \citet{ren2021portaspeech}, we evaluate the prosody from the aspects of pitch and duration: 1) we compute the average dynamic time warping (DTW) \cite{muller2007dynamic} distances between the pitch contours of GT speech and synthesized speech to measure the pitch accuracy; 2) we calculate the average absolute duration error (DE) in micro-seconds\footnote{In our PB/VB TTS baseline, the duration is predicted in phoneme/word level, respectively.} to measure the duration accuracy.
	
	\subsection{Performance}
	We compare the performance of our CLAPSpeech against BERT and A$^3$T in PB/VB TTS models. GT (the ground-truth audio) and GT (voc.) (the audio waveform generated by the vocoder using the GT mel-spectrogram) are also included in the experiment. We perform the TTS experiments on three datasets as mentioned in Section \ref{sec:experimental_setup}. The results are shown in Table \ref{tab:main_experiment}. We can see that CLAPSpeech outperforms other representation learning methods in both PB and VB TTS baselines in terms of MOS, pitch accuracy, and duration accuracy, which proves that CLAPSpeech could effectively improve the prosody prediction in current expressive TTS models (no matter prediction-based or variation-based). Besides, we observe that CLAPSpeech achieves better performance than BERT and A$^3$T with much fewer model parameters. We suspect it is due to the fact that the MLM-based method (i.e., BERT) require a large model capacity to store the semantic information and MAM-based method (i.e., A$^3$T) have to jointly learn the phonetic information to reconstruct the masked mel-spectrogram. By contrast, our CLAPSpeech eliminates the phonetic space and only focus on the prosody space during pre-training, which is parameter-efficient.
	
\begin{figure*}
		\small
		\centering
		\subfigure[GT]{
			\begin{minipage}[t]{0.18\linewidth}
				\centering
				\includegraphics[width=0.99\linewidth]{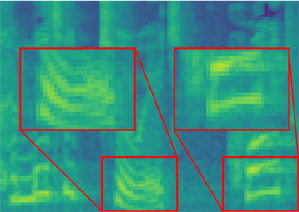}
			\end{minipage}
			\label{fig:mel_gt}
		}%
		\subfigure[PB]{
			\begin{minipage}[t]{0.18\linewidth}
				\centering
				\includegraphics[width=0.99\linewidth]{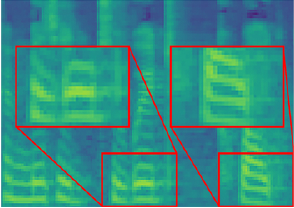}
			\end{minipage}
			\label{fig:mel_fs}
		}
		\subfigure[PB+BERT]{
			\begin{minipage}[t]{0.188\linewidth}
				\centering
				\includegraphics[width=0.99\linewidth]{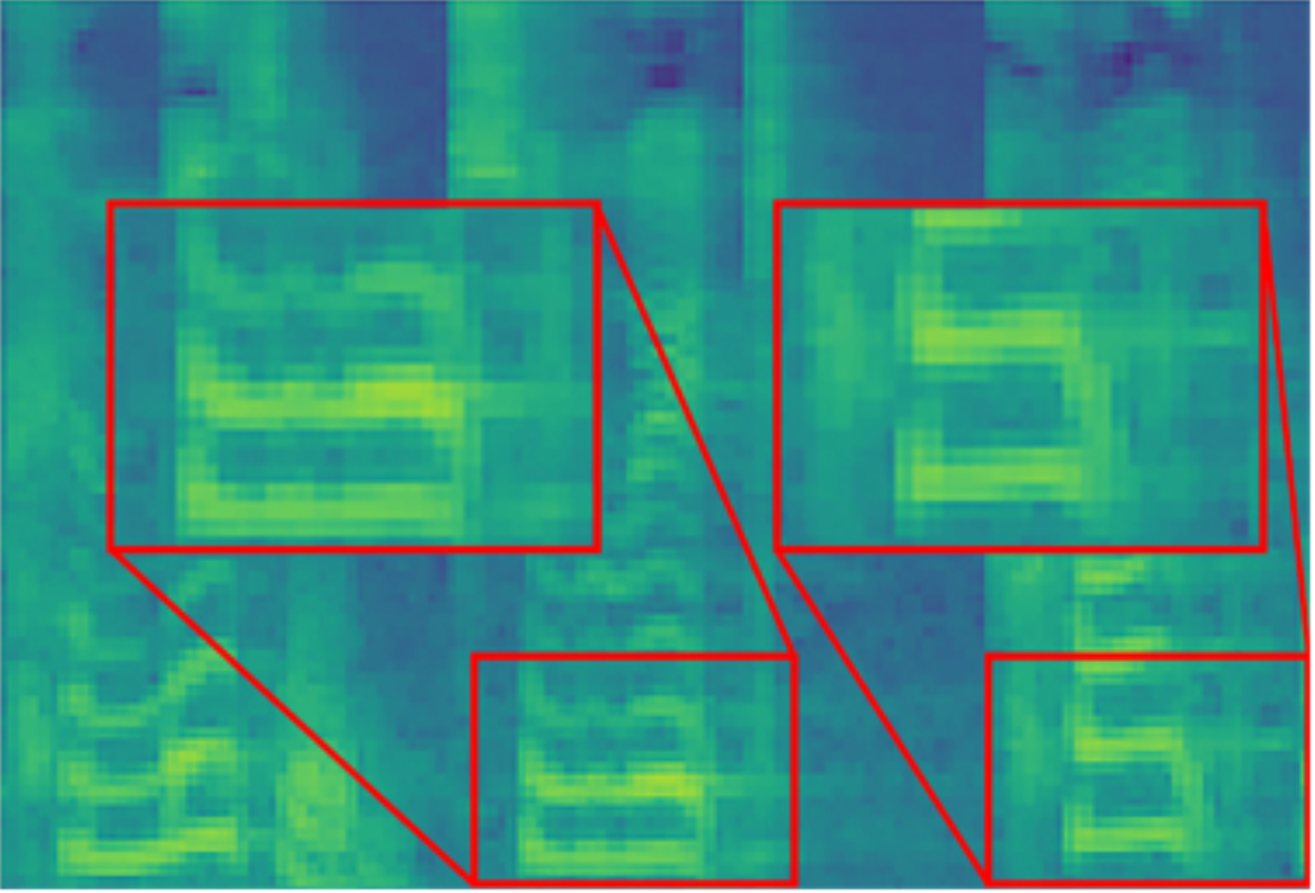}
			\end{minipage}
			\label{fig:mel_bert}
		}%
		\subfigure[PB+A$^3$T]{
			\begin{minipage}[t]{0.178\linewidth}
				\centering
				\includegraphics[width=0.99\linewidth]{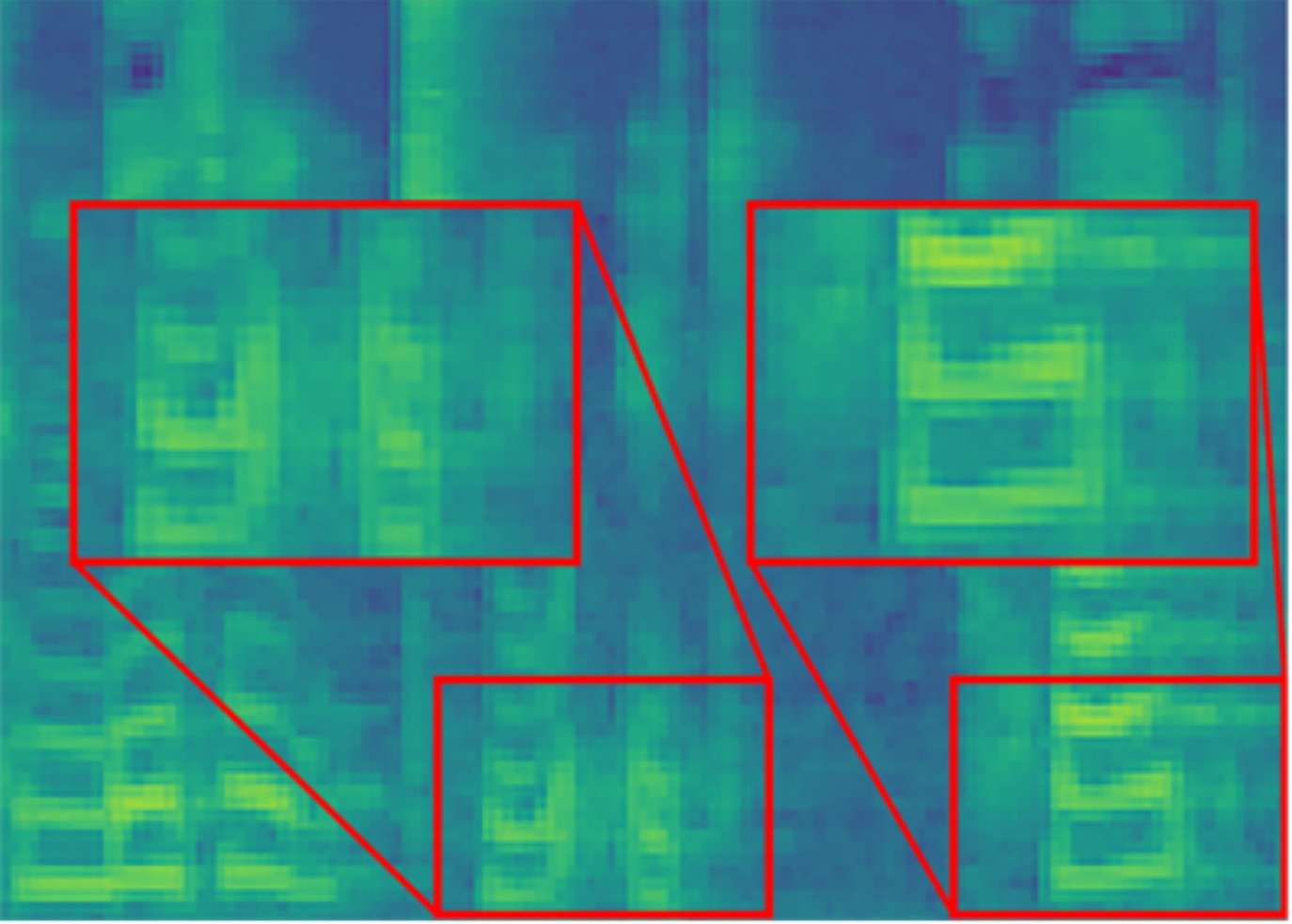}
			\end{minipage}
			\label{fig:mel_a3t}
		}
		\subfigure[PB + CLAPSpeech]{
			\begin{minipage}[t]{0.184\linewidth}
				\centering
				\includegraphics[width=0.99\linewidth]{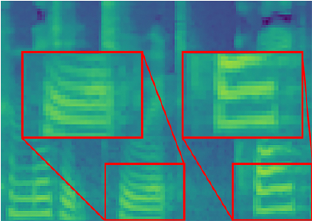}
			\end{minipage}
			\label{fig:mel_clip}
		}
		\caption{Visualizations of the mel-spectrograms generated by different TTS systems.	}
		\label{figure:mels}	
	\end{figure*}

	\begin{table*}[t]
		\caption{Performance comparison of different methods. \textit{PB} and \textit{VB} denote \textit{prediction-based} and \textit{variaition-based} TTS baselines, respectively. DTW denotes the dynamic time warping distance of pitch contours in the Mel-spectrogram. DE means the averaged absolute duration error in micro-seconds.}
		\begin{center}
			\centering
			\small
				\begin{tabular}{l | ccc | ccc | ccc | c}
					\toprule
					{\multirow{2}{*}{Method}} & \multicolumn{3}{c}{LJSpeech} & \multicolumn{3}{c}{Biaobei} & \multicolumn{3}{c}{LibriTTS} & \multirow{2}{*}{\#Params}\\
					& MOS$\uparrow$& DTW$\downarrow$& DE$\downarrow$ & MOS$\uparrow$ & DTW$\downarrow$  & DE$\downarrow$  & MOS$\uparrow$ & DTW$\downarrow$ & DE$\downarrow$   &\\
					\midrule
					\textit{GT}      &  4.81 & 0 & 0 &            4.59 &  0 & 0&                4.40 & 0 & 0  & /\\
					\textit{GT(voc.) }     & 4.63 & 0 & 0 &           4.43 &  0 &  0 &                 4.26 & 0 & 0  & /\\
					\midrule
					\textit{PB}      &  3.77 & 29.09 & 25.77 &          3.37 &  18.01 &  28.79 &                 3.43 & 14.26 & 27.42  & 11.99M\\
					\textit{PB + BERT}     &  4.04 & 27.43 & 24.97 &         3.43   &  16.79 &  28.06 &                 3.60 & 13.82 & 26.70  & 109.48M\\
					\textit{PB + A$^3$T }     &  3.92 & 28.18 & 25.63 &           3.51 &  17.18 &  28.44 &                 3.54 & 13.67 & 27.03 & 48.25M\\
					\textit{PB + CLAPSpeech}     &  \textbf{4.11} & \textbf{27.16} & \textbf{24.19} &           \textbf{3.62} &  \textbf{16.04} & \textbf{27.60} &                			\textbf{3.71} & \textbf{13.37} & \textbf{26.46} & 30.51M\\
					\midrule
					\textit{VB}      &  3.96 & 27.58 & 53.23 &          3.75  &  14.22 &  40.31 &                 3.81 & 11.96 & 52.51 & 23.02M\\
					\textit{VB + BERT}     &  4.13 & 26.97 & 52.01 &         3.91   &  13.63 & 38.41 &                 3.95 & 11.51 & 51.27  & 132.69M\\
					\textit{VB + A$^3$T }     &  4.05 & 26.37 & 52.17 &          4.04  &  13.97 &  39.15 &                 3.82 &11.71 & 51.98 & 59.73M\\
					\textit{VB + CLAPSpeech}     &  \textbf{4.28} & \textbf{25.94} & \textbf{51.34} &           \textbf{4.22} & \textbf{13.48} & \textbf{37.07} &                 		\textbf{4.06} & \textbf{10.93} & \textbf{50.89} & 41.54M\\
					\bottomrule
				\end{tabular}
		\end{center}
		\label{tab:main_experiment}
	\end{table*}
	We then visualize the mel-spectrograms generated by different methods in Figure \ref{figure:mels}. We can see that CLAPSpeech can generate results with more realistic pitch contours, which result in expressive prosody. In conclusion, our experiments demonstrate that CLAPSpeech could help TTS systems synthesize more expressive and prosodic audio.

	\subsection{Deeper Analysis}
	\label{sec:deeper_analysis}
	\subsubsection{Token Representation Self-similarity}
	\label{sec:discuss_self_similarity}
	To better understand the performance superiority of CLAPSPeech over existing representation learning methods for TTS, we analyze the token representation learned by CLAPSpeech and other methods. Following \citet{su2021tacl}, we define the averaged similarity on the selected token under different contexts $T=[T_1, ..., T_N]$ as, 
	\begin{equation}
		s(T) = \frac{1}{N(N-1)}\sum_{i=1}^{N}\sum_{j=1,j\neq i}^{N} cosine(T_i, T_j)
	\end{equation}
	where $T_i$ and $T_j$ are the selected token's encoding extracted by the model from different text contexts. Intuitively, a lower $s(T)$ indicates that the selected token itself plays a smaller role in generating its representation, which means that the model captures more context-related information from the input text sequence, and thus predicts better prosody.
	
	\paragraph{Quantitative Evaluation}
	We sample 10,000 batches (each batch consists of 256 sentences that contain the same selected token) from the ASR validation datasets and compute the averaged self-similarity. The result is shown in Table \ref{tab:self_similarity}. We observe that our CLAPSpeech learned with the contrastive objective (in Equation \ref{eq:clip_ph}) achieves the lowest similarity in the off-diagonal entries of the similarity matrix, which denotes that the model has made use of the text context to capture the prosody variance of the same token, thus achieve the best prosody performance in Table \ref{tab:main_experiment}. Besides, we can see that BERT also achieves a relatively low off-diagonal similarity, which is due to its MLM task during pre-training, in which the model needs to extract semantic information from context to predict the masked token. By contrast, the vanilla TTS text encoder and A$^3$T fail to achieve a low off-diagonal similarity, which means that both models cannot extract discriminative information from different contexts. We suspect the failure of A$^3$T is due to the fact that its MAM objective encourages the model to predict the masked mel-spectrogram patch based on the input unmasked text sequence, which increases the model's demand for phonetic information of the selected token. 
	
	\begin{table}[t]
		\caption{Self-similarity score of different methods. TTS denotes the text encoder of the vanilla TTS baseline.}
		\begin{center}
			\centering
			\resizebox{1.0\columnwidth}{!}{%
				\begin{tabular}{lcccc}
					\toprule
					\textit{Text Encoder of } &TTS & BERT &  A$^3$T &  CLAPSPeech \\
					\midrule
					\textbf{Self-Similarity} &  0.9854 & 0.5517 &0.9390 &\textbf{ 0.4160 }\\
					\bottomrule
				\end{tabular}
			}
		\end{center}
		\label{tab:self_similarity}
	\end{table}
	
\begin{figure}
		\centering
		\subfigure[CLAPSpeech]{
			\begin{minipage}[t]{0.475\linewidth}
				\centering
				\includegraphics[width=0.99\linewidth]{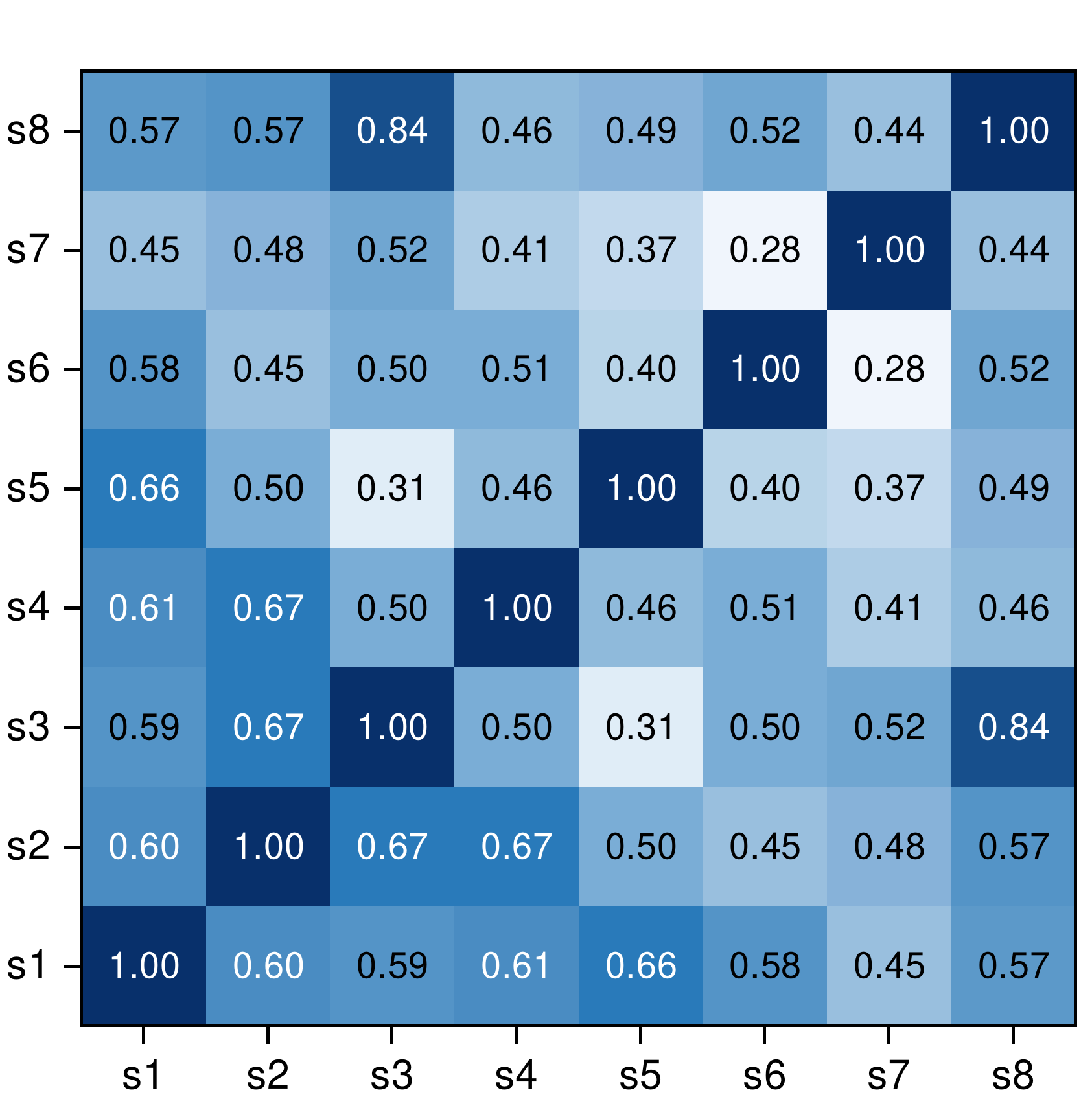}
			\end{minipage}
			\label{fig:heatmap_clip}
		}%
		\subfigure[TTS]{
			\begin{minipage}[t]{0.51\linewidth}
				\centering
				\includegraphics[width=0.99\linewidth]{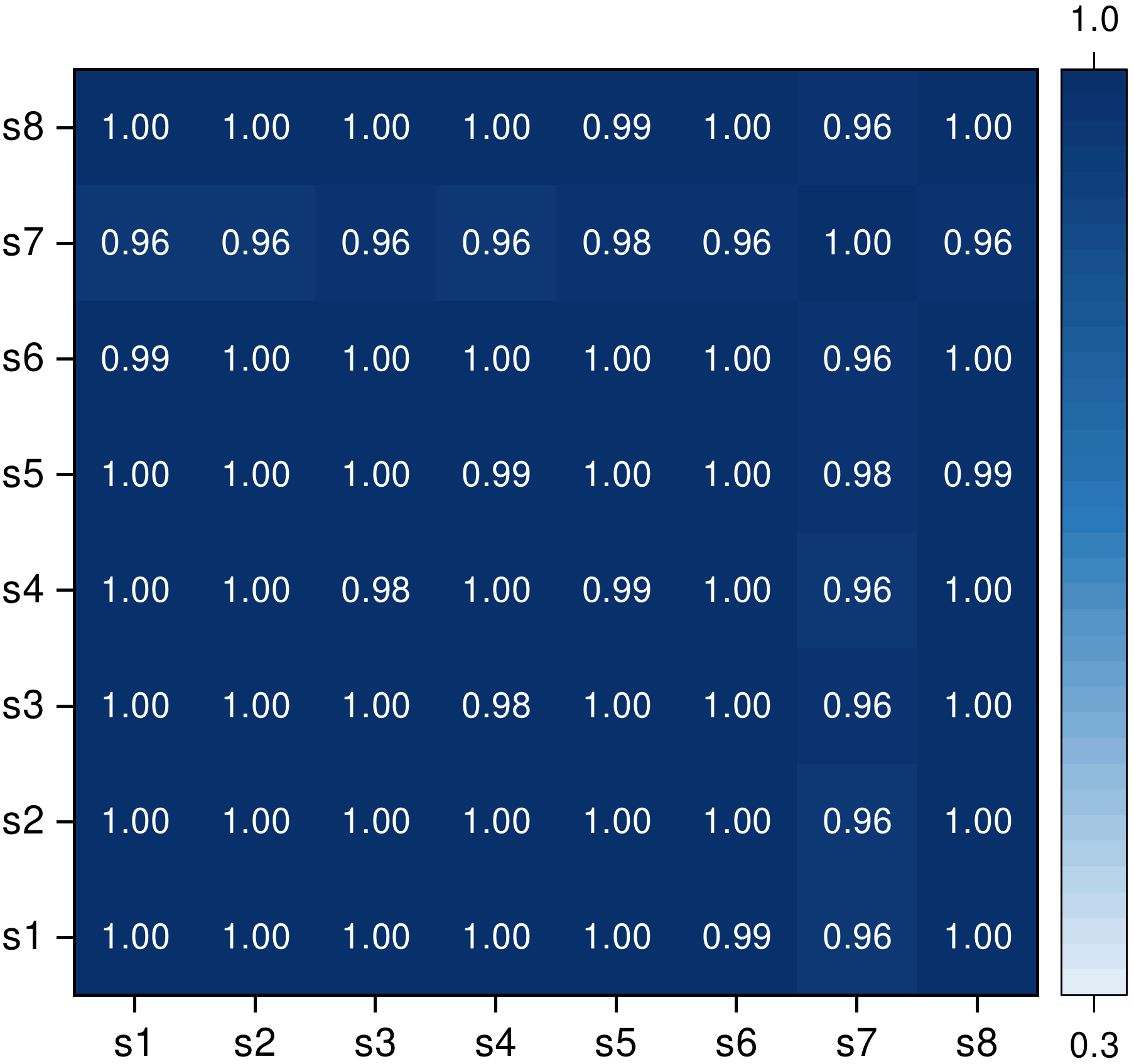}
			\end{minipage}
			\label{fig:heatmap_tst}
		}
	
		\caption{Example: self-similarity matrix visualization of CLAPSpeech and the text encoder of the vanilla TTS model. $s_i$ denotes the $i_{th}$ sentence.}
		\label{fig:heatmap}
	\end{figure}

	\paragraph{Qualitative Evaluation}
	We sample 8 sentences\footnote{We list these sentences in Table \ref{tab:8_sentences} of Appendix \ref{appendix:analysis}.} that contain the word \textit{"higher"} from LibriSpeech and visualize the self-similarity matrix $M$ (where $M_{i,j} = cosine(T_i, T_j)$) produced by CLAPSpeech and vanilla TTS text encoder. The results are shown in Figure \ref{fig:heatmap}, where a darker color denotes a higher self-similarity score. We also provide the self-similarity matrix of BERT and A$^3$T in Figure \ref{fig:heatmap2} of Appendix \ref{appendix:analysis}. We can see that the self-similarities of CLAPSpeech are much lower in the off-diagonal entries.
	
	\subsubsection{Fine-grained Prosody Transfer}
	\label{sec:prosody_transfer}
\begin{figure}
		\centering
		\subfigure[reference (s7)]{
			\begin{minipage}[t]{0.3\linewidth}
				\centering
				\includegraphics[width=0.99\linewidth]{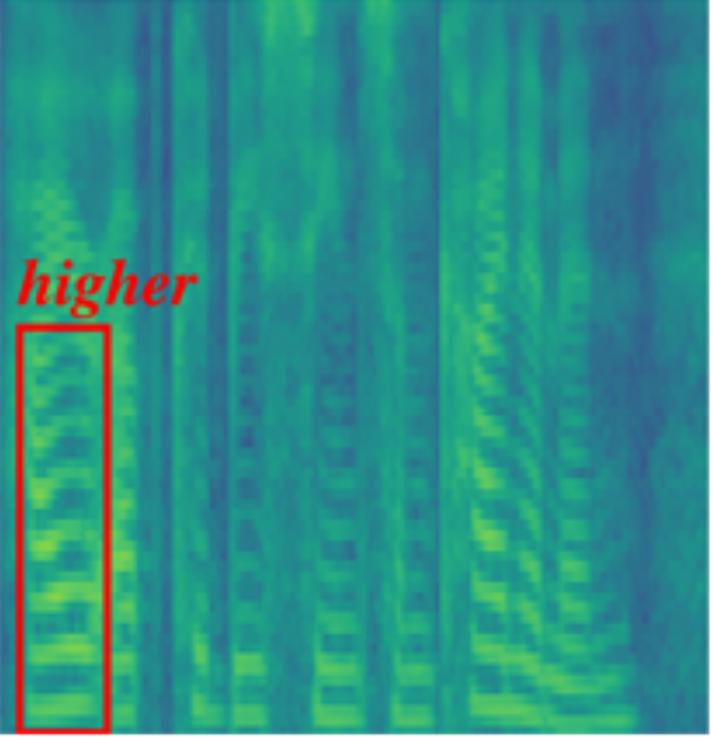}
			\end{minipage}
			\label{fig:mel_ref}
		}%
		\subfigure[source (s8)]{
			\begin{minipage}[t]{0.3\linewidth}
				\centering
				\includegraphics[width=0.99\linewidth]{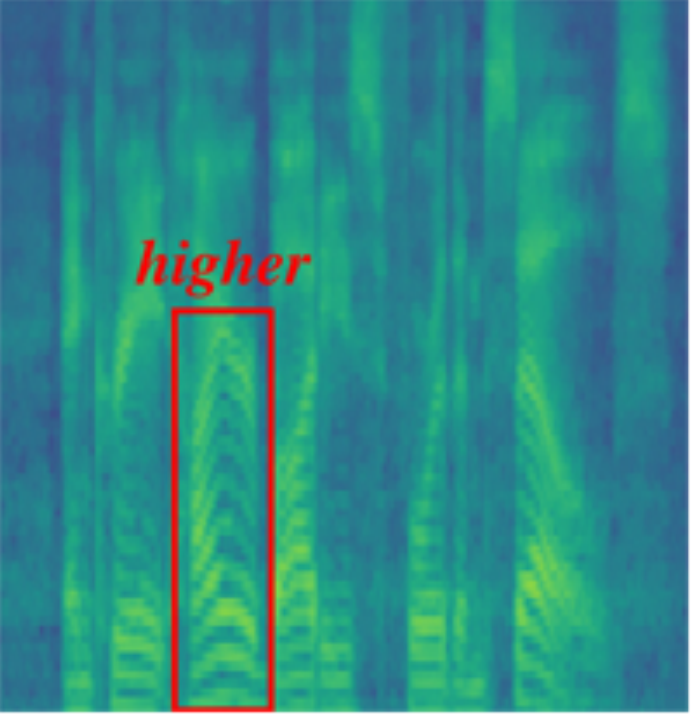}
			\end{minipage}
			\label{fig:mel_before}
		}
		\subfigure[transferred (s8)]{
			\begin{minipage}[t]{0.3\linewidth}
				\centering
				\includegraphics[width=0.99\linewidth]{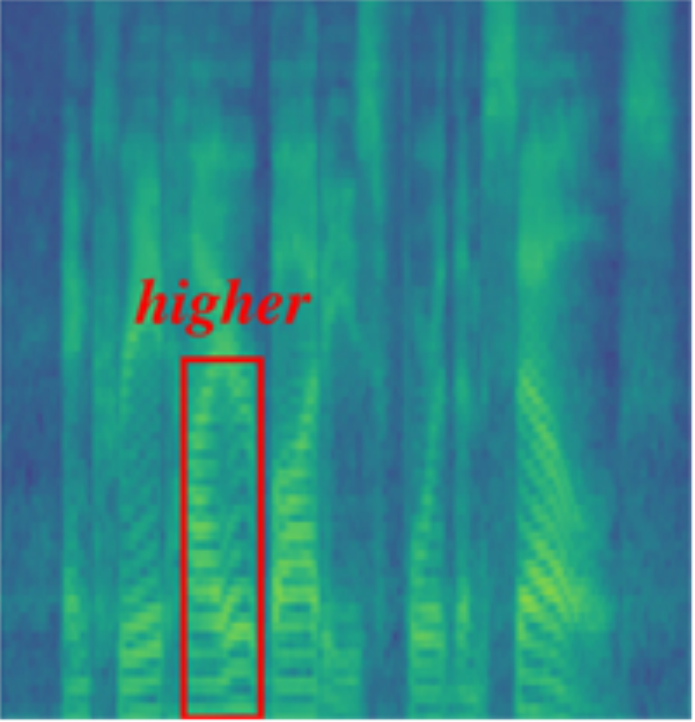}
			\end{minipage}
			\label{fig:mel_after}
		}%
		\caption{Visualizations of the mel-spectrograms generated in prosody transfer.	}
		\label{figure:mel_transfer}	
	\end{figure}

	We perform an intuitive case study about prosody transfer to further validate that our CLAPSpeech's text-speech joint multi-modal space represents high-level prosody patterns (i.e., the pitch contours and duration information). We take s7/8 in Table \ref{tab:8_sentences} as the reference/source audio and expect to transfer the word \textit{"higher"}'s prosody pattern from s7 to s8. Specifically, we use the text encoder of CLAPSpeech to extract the text prosody encoding of s7 and s8, then replace the text token encoding of \textit{"higher"} in s8 with that in s7. As shown in Figure \ref{figure:mel_transfer}, the prosody pattern of \textit{"higher"} in s8\footnote{the pitch contours in reference remain flat in the early stage and then rise in the late stage} in Figure \ref{fig:mel_ref}  has been successfully transferred into s7 in Figure \ref{fig:mel_after}. We also provide audio samples of this case study on our demo page. The manipulation of the local prosody proves that our CLAPSpeech extract prosody representation effectively influences the prosody prediction of the TTS system.

	\subsection{Ablation Studies}
	\paragraph{Use BPE as Auxiliary Features} We first analyze the effectiveness of the BPE as an auxiliary feature to help extract prosody information from the text context. During the pre-training phase of CLAPSpeech, we found removing BPE from the text encoder significantly degrades the validation CLIP loss from 0.3692 to 0.6764. Then in the TTS phase, as can be seen in line 3 in Table \ref{tab:ablation}, the ablated model using the pre-trained text encoder without BPE leads to a performance drop in terms of CMOS, DTW, and DE. This is possibly due to the fact that BPE could better represent the semantic information than the low-level phoneme sequence.
	
	\paragraph{Multi-scale Pre-training} To demonstrate the effectiveness of multi-scale pre-training, as can be seen in line 4/5 in Table \ref{tab:ablation}, we tried to remove phoneme-level or word-level CLAPSpeech from the model, which leads to a worse prosody performance. We also tried to use the untrained CLAPSpeech to prove the necessity of the pre-training process, and we found this ablated model (line 6) achieves a slightly worse performance than the TTS baseline (line 3).
	
	\begin{table}[!htb]
		\small
		\centering
		\caption{Performance comparison for ablation studies.}
		\begin{tabular}{ l | c| c | c }
			\toprule
			Setting & CMOS & DTW & DE \\
			
			\midrule
			\textit{TTS + CLAPSpeech} & \textbf{0} & \textbf{27.16} & \textbf{24.19} \\
			\textit{TTS baseline} &  -1.53 & 29.09 & 25.77 \\
			\midrule
			\textit{w/o BPE} & -1.08 & 28.21 & 24.93 \\ %
			\textit{w/o ph-level} & -1.11 & 27.68 & 25.01 \\ %
			\textit{w/o word-level} & -0.46 & 27.55 & 24.52 \\
			\textit{untrained} & -1.67 &29.45 & 25.96\\ %
			\bottomrule
		\end{tabular}
		
		\label{tab:ablation}
	\end{table}
	
	\section{Conclusion}
	In this paper, we propose CLAPSpeech, a cross-modal contrastive pre-training framework that provides better text representation with rich prosody information for TTS. With the design of a  text encoder and a prosody encoder, CLAPSpeech learns to connect the text context with its corresponding prosody pattern in the speech. We also introduced multi-scale pre-training to extract prosody patterns at multiple levels. We have demonstrated the performance and generalization ability of CLAPSpeech on three TTS datasets (English, Chinese, and multi-speaker, respectively). We have also deeply analyzed the principle behind the improvement of CLAPSpeech and performed ablation studies to prove the necessity of each component. 
	
	\section{Limitations}
	There are majorly two limitations: Firstly, in this work, we only consider the current-sentence text context-related prosody. In future work, we will focus on improving the inter-sentence prosody to achieve coherent, expressive TTS for long-form text. Secondly, other variables are not considered during the contrastive pre-training. One can explore similar approaches that connect prosody to other conditions such as speaker, emotion, etc. 
	
	\section{Ethics Statement}
	CLAPSpeech improves the prosody of the synthesized speech, which may cause unemployment for people with related occupations. Besides, the production of fake speeches may cause voice security issues. Further efforts in automatic speaker verification should be made to improve voice security.
	
	\section{Acknowledgment}
	This work was supported in part by the National Key R\&D Program of China under Grant No.2022ZD0162000,National Natural Science Foundation of China under Grant No. 62222211 and Grant No.61836002 and Grant No.62072397, and Yiwise.
	
	\bibliography{custom}

\begin{thebibliography}{38}
\expandafter\ifx\csname natexlab\endcsname\relax\def\natexlab#1{#1}\fi

\bibitem[{Baevski et~al.(2020)Baevski, Zhou, Mohamed, and
  Auli}]{baevski2020wav2vec}
Alexei Baevski, Yuhao Zhou, Abdelrahman Mohamed, and Michael Auli. 2020.
\newblock wav2vec 2.0: A framework for self-supervised learning of speech
  representations.
\newblock In \emph{NIPS}.

\bibitem[{Bai et~al.(2022)Bai, Zheng, Chen, Ma, Li, and Huang}]{bai2022a3t}
He~Bai, Renjie Zheng, Junkun Chen, Mingbo Ma, Xintong Li, and Liang Huang.
  2022.
\newblock A3t: Alignment-aware acoustic and text pretraining for speech
  synthesis and editing.
\newblock In \emph{ICML}.

\bibitem[{Chen et~al.(2020)Chen, Ma, Zheng, and Huang}]{chen2020mam}
Junkun Chen, Mingbo Ma, Renjie Zheng, and Liang Huang. 2020.
\newblock Mam: Masked acoustic modeling for end-to-end speech-to-text
  translation.
\newblock \emph{arXiv preprint arXiv:2010.11445}.

\bibitem[{Chen et~al.(2021)Chen, Deng, Wang, Soong, and He}]{chen2021speech}
Liping Chen, Yan Deng, Xi~Wang, Frank~K Soong, and Lei He. 2021.
\newblock Speech bert embedding for improving prosody in neural tts.
\newblock In \emph{ICASSP}.

\bibitem[{Chung et~al.(2019)Chung, Wang, Hsu, Zhang, and
  Skerry-Ryan}]{chung2019semi}
Yu-An Chung, Yuxuan Wang, Wei-Ning Hsu, Yu~Zhang, and RJ~Skerry-Ryan. 2019.
\newblock Semi-supervised training for improving data efficiency in end-to-end
  speech synthesis.
\newblock In \emph{ICASSP}.

\bibitem[{Devlin et~al.(2019)Devlin, Chang, Lee, and
  Toutanova}]{devlin2019bert}
Jacob Devlin, Ming-Wei Chang, Kenton Lee, and Kristina Toutanova. 2019.
\newblock Bert: Pre-training of deep bidirectional transformers for language
  understanding.
\newblock In \emph{NAACL-HLT}.

\bibitem[{Donahue et~al.(2021)Donahue, Dieleman, Bi{\'n}kowski, Elsen, and
  Simonyan}]{donahue2020end}
Jeff Donahue, Sander Dieleman, Miko{\l}aj Bi{\'n}kowski, Erich Elsen, and Karen
  Simonyan. 2021.
\newblock End-to-end adversarial text-to-speech.
\newblock In \emph{ICLR}.

\bibitem[{Elias et~al.(2021)Elias, Zen, Shen, Zhang, Jia, Weiss, and
  Wu}]{elias2021parallel}
Isaac Elias, Heiga Zen, Jonathan Shen, Yu~Zhang, Ye~Jia, Ron~J Weiss, and
  Yonghui Wu. 2021.
\newblock Parallel tacotron: Non-autoregressive and controllable tts.
\newblock In \emph{ICASSP}.

\bibitem[{Elizalde et~al.(2022)Elizalde, Deshmukh, Ismail, and
  Wang}]{elizalde2022clap}
Benjamin Elizalde, Soham Deshmukh, Mahmoud~Al Ismail, and Huaming Wang. 2022.
\newblock Clap: Learning audio concepts from natural language supervision.
\newblock \emph{arXiv preprint arXiv:2206.04769}.

\bibitem[{He et~al.(2016)He, Zhang, Ren, and Sun}]{he2016deep}
Kaiming He, Xiangyu Zhang, Shaoqing Ren, and Jian Sun. 2016.
\newblock Deep residual learning for image recognition.
\newblock In \emph{CVPR}.

\bibitem[{Hsu et~al.(2021)Hsu, Bolte, Tsai, Lakhotia, Salakhutdinov, and
  Mohamed}]{hsu2021hubert}
Wei-Ning Hsu, Benjamin Bolte, Yao-Hung~Hubert Tsai, Kushal Lakhotia, Ruslan
  Salakhutdinov, and Abdelrahman Mohamed. 2021.
\newblock Hubert: Self-supervised speech representation learning by masked
  prediction of hidden units.
\newblock \emph{IEEE/ACM Transactions on Audio, Speech, and Language
  Processing}, 29:3451--3460.

\bibitem[{Ito and Johnson(2017)}]{ljspeech17}
Keith Ito and Linda Johnson. 2017.
\newblock The lj speech dataset.
\newblock \url{https://keithito.com/LJ-Speech-Dataset/}.

\bibitem[{Jia et~al.(2021)Jia, Zen, Shen, Zhang, and Wu}]{jia2021png}
Ye~Jia, Heiga Zen, Jonathan Shen, Yu~Zhang, and Yonghui Wu. 2021.
\newblock Png bert: Augmented bert on phonemes and graphemes for neural tts.

\bibitem[{Jia et~al.(2018)Jia, Zhang, Weiss, Wang, Shen, Ren, Nguyen, Pang,
  Lopez~Moreno, Wu et~al.}]{jia2018transfer}
Ye~Jia, Yu~Zhang, Ron Weiss, Quan Wang, Jonathan Shen, Fei Ren, Patrick Nguyen,
  Ruoming Pang, Ignacio Lopez~Moreno, Yonghui Wu, et~al. 2018.
\newblock Transfer learning from speaker verification to multispeaker
  text-to-speech synthesis.
\newblock \emph{NIPS}.

\bibitem[{Jiang et~al.(2022)Jiang, Zhe, Zhao, Yang, Ren, Liu, and
  Ye}]{jiang2022dict}
Ziyue Jiang, Su~Zhe, Zhou Zhao, Qian Yang, Yi~Ren, Jinglin Liu, and Zhenhui Ye.
  2022.
\newblock Dict-tts: Learning to pronounce with prior dictionary knowledge for
  text-to-speech.
\newblock \emph{arXiv preprint arXiv:2206.02147}.

\bibitem[{Kim et~al.(2020)Kim, Kim, Kong, and Yoon}]{kim2020glow}
Jaehyeon Kim, Sungwon Kim, Jungil Kong, and Sungroh Yoon. 2020.
\newblock Glow-tts: A generative flow for text-to-speech via monotonic
  alignment search.
\newblock In \emph{NIPS}.

\bibitem[{Kim et~al.(2021)Kim, Kong, and Son}]{kim2021conditional}
Jaehyeon Kim, Jungil Kong, and Juhee Son. 2021.
\newblock Conditional variational autoencoder with adversarial learning for
  end-to-end text-to-speech.
\newblock In \emph{ICML}.

\bibitem[{Kong et~al.(2020)Kong, Kim, and Bae}]{kong2020hifi}
Jungil Kong, Jaehyeon Kim, and Jaekyoung Bae. 2020.
\newblock Hifi-gan: Generative adversarial networks for efficient and high
  fidelity speech synthesis.
\newblock In \emph{NIPS}.

\bibitem[{Liu et~al.(2022)Liu, Li, Ren, Chen, and Zhao}]{liu2022diffsinger}
Jinglin Liu, Chengxi Li, Yi~Ren, Feiyang Chen, and Zhou Zhao. 2022.
\newblock Diffsinger: Singing voice synthesis via shallow diffusion mechanism.
\newblock In \emph{AAAI}.

\bibitem[{Liu et~al.(2021)Liu, Sisman, and Li}]{liu2021graphspeech}
Rui Liu, Berrak Sisman, and Haizhou Li. 2021.
\newblock Graphspeech: Syntax-aware graph attention network for neural speech
  synthesis.
\newblock In \emph{ICASSP}.

\bibitem[{Miao et~al.(2021)Miao, Shuang, Liu, Minchuan, Ma, Wang, and
  Xiao}]{miao2021efficienttts}
Chenfeng Miao, Liang Shuang, Zhengchen Liu, Chen Minchuan, Jun Ma, Shaojun
  Wang, and Jing Xiao. 2021.
\newblock Efficienttts: An efficient and high-quality text-to-speech
  architecture.
\newblock In \emph{ICML}.

\bibitem[{Muller(2007)}]{muller2007dynamic}
Meinard Muller. 2007.
\newblock Dynamic time warping.
\newblock \emph{Information retrieval for music and motion}, pages 69--84.

\bibitem[{Panayotov et~al.(2015)Panayotov, Chen, Povey, and
  Khudanpur}]{panayotov2015librispeech}
Vassil Panayotov, Guoguo Chen, Daniel Povey, and Sanjeev Khudanpur. 2015.
\newblock Librispeech: an asr corpus based on public domain audio books.
\newblock In \emph{ICASSP}.

\bibitem[{Radford et~al.(2021)Radford, Kim, Hallacy, Ramesh, Goh, Agarwal,
  Sastry, Askell, Mishkin, Clark et~al.}]{radford2021clip}
Alec Radford, Jong~Wook Kim, Chris Hallacy, Aditya Ramesh, Gabriel Goh,
  Sandhini Agarwal, Girish Sastry, Amanda Askell, Pamela Mishkin, Jack Clark,
  et~al. 2021.
\newblock Learning transferable visual models from natural language
  supervision.
\newblock In \emph{ICML}.

\bibitem[{Ren et~al.(2021{\natexlab{a}})Ren, Hu, Tan, Qin, Zhao, Zhao, and
  Liu}]{ren2021fastspeech2}
Yi~Ren, Chenxu Hu, Xu~Tan, Tao Qin, Sheng Zhao, Zhou Zhao, and Tie-Yan Liu.
  2021{\natexlab{a}}.
\newblock Fastspeech 2: Fast and high-quality end-to-end text to speech.
\newblock In \emph{ICLR}.

\bibitem[{Ren et~al.(2022)Ren, Lei, Huang, Zhang, Chen, Yan, and
  Zhao}]{ren2022prosospeech}
Yi~Ren, Ming Lei, Zhiying Huang, Shiliang Zhang, Qian Chen, Zhijie Yan, and
  Zhou Zhao. 2022.
\newblock Prosospeech: Enhancing prosody with quantized vector pre-training in
  text-to-speech.
\newblock In \emph{ICASSP}.

\bibitem[{Ren et~al.(2021{\natexlab{b}})Ren, Liu, and
  Zhao}]{ren2021portaspeech}
Yi~Ren, Jinglin Liu, and Zhou Zhao. 2021{\natexlab{b}}.
\newblock Portaspeech: Portable and high-quality generative text-to-speech.
\newblock In \emph{NIPS}.

\bibitem[{Ren et~al.(2019)Ren, Ruan, Tan, Qin, Zhao, Zhao, and
  Liu}]{ren2019fastspeech}
Yi~Ren, Yangjun Ruan, Xu~Tan, Tao Qin, Sheng Zhao, Zhou Zhao, and Tie-Yan Liu.
  2019.
\newblock Fastspeech: Fast, robust and controllable text to speech.

\bibitem[{Shibata et~al.(1999)Shibata, Kida, Fukamachi, Takeda, Shinohara,
  Shinohara, and Arikawa}]{shibata1999byte}
Yusuxke Shibata, Takuya Kida, Shuichi Fukamachi, Masayuki Takeda, Ayumi
  Shinohara, Takeshi Shinohara, and Setsuo Arikawa. 1999.
\newblock Byte pair encoding: A text compression scheme that accelerates
  pattern matching.

\bibitem[{Su et~al.(2021)Su, Liu, Meng, Lan, Shu, Shareghi, and
  Collier}]{su2021tacl}
Yixuan Su, Fangyu Liu, Zaiqiao Meng, Tian Lan, Lei Shu, Ehsan Shareghi, and
  Nigel Collier. 2021.
\newblock Tacl: Improving bert pre-training with token-aware contrastive
  learning.
\newblock \emph{arXiv preprint arXiv:2111.04198}.

\bibitem[{Tan et~al.(2021)Tan, Qin, Soong, and Liu}]{tan2021survey}
Xu~Tan, Tao Qin, Frank Soong, and Tie-Yan Liu. 2021.
\newblock A survey on neural speech synthesis.
\newblock \emph{arXiv preprint arXiv:2106.15561}.

\bibitem[{Vaswani et~al.(2017)Vaswani, Shazeer, Parmar, Uszkoreit, Jones,
  Gomez, Kaiser, and Polosukhin}]{vaswani2017attention}
Ashish Vaswani, Noam Shazeer, Niki Parmar, Jakob Uszkoreit, Llion Jones,
  Aidan~N Gomez, {\L}ukasz Kaiser, and Illia Polosukhin. 2017.
\newblock Attention is all you need.
\newblock In \emph{NIPS}.

\bibitem[{Wang et~al.(2015)Wang, Qian, Soong, He, and Zhao}]{wang2015word}
Peilu Wang, Yao Qian, Frank~K Soong, Lei He, and Hai Zhao. 2015.
\newblock Word embedding for recurrent neural network based tts synthesis.
\newblock In \emph{ICASSP}.

\bibitem[{Wang et~al.(2018)Wang, Stanton, Zhang, Ryan, Battenberg, Shor, Xiao,
  Jia, Ren, and Saurous}]{wang2018style}
Yuxuan Wang, Daisy Stanton, Yu~Zhang, RJ-Skerry Ryan, Eric Battenberg, Joel
  Shor, Ying Xiao, Ye~Jia, Fei Ren, and Rif~A Saurous. 2018.
\newblock Style tokens: Unsupervised style modeling, control and transfer in
  end-to-end speech synthesis.
\newblock In \emph{ICML}.

\bibitem[{Ye et~al.(2022)Ye, Zhao, Ren, and Wu}]{ye2022syntaspeech}
Zhenhui Ye, Zhou Zhao, Yi~Ren, and Fei Wu. 2022.
\newblock Syntaspeech: Syntax-aware generative adversarial text-to-speech.
\newblock \emph{arXiv preprint arXiv:2204.11792}.

\bibitem[{Zen et~al.(2019)Zen, Dang, Clark, Zhang, Weiss, Jia, Chen, and
  Wu}]{zen2019libritts}
Heiga Zen, Viet Dang, Rob Clark, Yu~Zhang, Ron~J Weiss, Ye~Jia, Zhifeng Chen,
  and Yonghui Wu. 2019.
\newblock Libritts: A corpus derived from librispeech for text-to-speech.
\newblock \emph{arXiv preprint arXiv:1904.02882}.

\bibitem[{Zhang et~al.(2022)Zhang, Lv, Guo, Shao, Yang, Xie, Xu, Bu, Chen, Zeng
  et~al.}]{zhang2022wenetspeech}
Binbin Zhang, Hang Lv, Pengcheng Guo, Qijie Shao, Chao Yang, Lei Xie, Xin Xu,
  Hui Bu, Xiaoyu Chen, Chenchen Zeng, et~al. 2022.
\newblock Wenetspeech: A 10000+ hours multi-domain mandarin corpus for speech
  recognition.
\newblock In \emph{ICASSP}.

\bibitem[{Zhang et~al.(2019)Zhang, Wang, Fang, Li, and
  Yamagishi}]{zhang2019joint}
Mingyang Zhang, Xin Wang, Fuming Fang, Haizhou Li, and Junichi Yamagishi. 2019.
\newblock Joint training framework for text-to-speech and voice conversion
  using multi-source tacotron and wavenet.
\newblock In \emph{INTERSPEECH}.

\end{thebibliography}
	\bibliographystyle{acl_natbib}

	\appendix
	\section{Details of Models}
	\label{appendix:details_of_models}

	\subsection{CLAPSpeech plugged in FastSpeech 2}
	\label{appendix:clapspeech_fs2}
\begin{figure}[!t]
		\centering
		\includegraphics[width=0.65\linewidth]{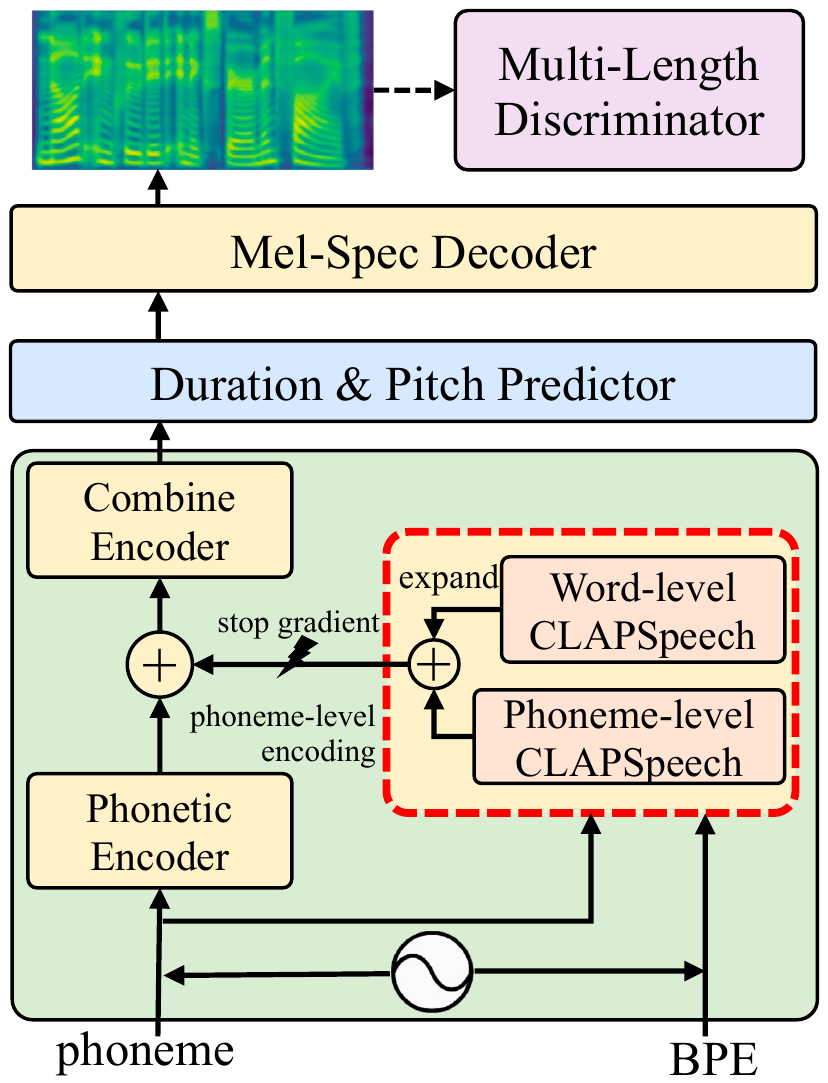}
		\caption{FastSpeech 2 with CLAPSpeech plugged in.}
		\vspace{-3mm}
		\label{fig:clapspeech_fs2_adv}
	\end{figure}

	We show how to integrate CLAPSpeech into a popular prediction-based TTS system, \textit{FastSpeech 2}. As shown in Figure \ref{fig:clapspeech_fs2_adv}, the pre-trained text encoders of CLAPSpeech (marked with a red dashed rectangle) perform as an auxiliary encoder to the original phonetic encoder of FastSpeech 2. The phoneme-level outputs of the phonetic encoder and CLAPSpeech text encoder are fused and processed by the following encoder. Note that we fix the parameters of CLAPSpeech text encoders during the training of the TTS system to avoid overfitting. 
	
	\subsection{Multi-length Adversarial Training}
	\label{appendix:multi-length}
	For the tested TTS baselines, we adopt an additional multi-length discriminator to provide a least squared GAN loss to improve the audio quality. The multi-length discriminator is an ensemble of multiple CNN-based discriminators which evaluates the mel-spectrogram based on random windows of different lengths. One could refer to \citet{ye2022syntaspeech} for more details.
	
	\section{Detailed Experimental Settings}
	\label{appendix:details_of_experiments}
	\subsection{Model Configurations}
	\label{appendix:hyper_params}
	We list the hyper-parameters of CLAPSpeech and the tested TTS baselines in Table \ref{tab:model_configs}.
	
	\begin{table*}[htbp]
		\small
		\centering
		\caption{The detailed model configurations.}
		\begin{tabular}{ccc|c}
			\hline
			\multicolumn{2}{c|}{Hyper-parameter} & CLAPSpeech & Number of parameters \bigstrut\\
			\hline
			\multicolumn{1}{c|}{\multirow{5}[2]{*}{Text Encoder}} & \multicolumn{1}{l|}{Phoneme/BPE embedding hidden size} & 192   & \multirow{5}[2]{*}{18.517M} \bigstrut[t]\\
			\multicolumn{1}{c|}{} & \multicolumn{1}{l|}{Phoneme/BPE encoder FFT blocks} & 4     &  \\
			\multicolumn{1}{c|}{} & \multicolumn{1}{l|}{Hidden size} & 192   &  \\
			\multicolumn{1}{c|}{} & \multicolumn{1}{l|}{Conv1D kernel} & 5     &  \\
			\multicolumn{1}{c|}{} & \multicolumn{1}{l|}{Conv1D filter size} & 768   &  \bigstrut[b]\\
			
			\hline
			\multicolumn{1}{c|}{\multirow{5}[2]{*}{Prosody Encoder}} & \multicolumn{1}{l|}{Residual blocks} & 4     & \multirow{6}[2]{*}{21.801M} \bigstrut[t]\\
			\multicolumn{1}{c|}{} & \multicolumn{1}{l|}{Number of conv layers per block} & 12   &  \\
			\multicolumn{1}{c|}{} & \multicolumn{1}{l|}{Hidden size} & 192     &  \\
			\multicolumn{1}{c|}{} & \multicolumn{1}{l|}{Input mel-spectrogram length} & 128   &  \\
			\multicolumn{1}{c|}{} & \multicolumn{1}{l|}{Hidden size in pooling layer} & 768     &  \\
			\multicolumn{1}{c|}{} & \multicolumn{1}{l|}{\#Attention heads in pooling layer} & 4     &  \bigstrut[b]\\
			\hline
			\multicolumn{1}{c|}{\multirow{4}[2]{*}{Prediction-based TTS baseline}} & \multicolumn{1}{l|}{Encoder Layers} & 4     & \multirow{4}[2]{*}{11.993M} \bigstrut[t]\\
			\multicolumn{1}{c|}{} & \multicolumn{1}{l|}{Decoder Layers} & 4     &  \\
			\multicolumn{1}{c|}{} & \multicolumn{1}{l|}{Encoder/Decoder Conv1D Kernel} & 9     &  \\
			\multicolumn{1}{c|}{} & \multicolumn{1}{l|}{Encoder/Decoder Conv1D channel size} & 256    \bigstrut[b]\\
			\hline
			\multicolumn{1}{c|}{\multirow{8}[2]{*}{Variation-based TTS baseline}} & \multicolumn{1}{l|}{Encoder Layers} & 8     & \multirow{8}[2]{*}{23.020M} \bigstrut[t]\\
			\multicolumn{1}{c|}{} & \multicolumn{1}{l|}{Decoder Layers} & 4     &  \\
			\multicolumn{1}{c|}{} & \multicolumn{1}{l|}{Encoder/Decoder Conv1D Kernel} & 5     &  \\
			\multicolumn{1}{c|}{} & \multicolumn{1}{l|}{Encoder/Decoder Conv1D channel size} & 192   &  \\
			\multicolumn{1}{c|}{} & \multicolumn{1}{l|}{Latent Size } & 16    &  \\
			\multicolumn{1}{c|}{} & \multicolumn{1}{l|}{Prior Flow Layers} & 4     &  \\
			\multicolumn{1}{c|}{} & \multicolumn{1}{l|}{Prior Flow Conv1D Kernel} & 3     &  \\
			\multicolumn{1}{c|}{} & \multicolumn{1}{l|}{Prior Flow Conv1D Channel Size} & 64    &  \bigstrut[b]\\
			\hline
			\multicolumn{1}{c|}{\multirow{4}[2]{*}{Multi-Length Discriminator}} & \multicolumn{1}{l|}{Number of CNN-based Discriminators} & 3     & \multirow{4}[2]{*}{0.927M} \bigstrut[t]\\
			\multicolumn{1}{c|}{} & \multicolumn{1}{l|}{Window size} & 32,64,128 &  \\
			\multicolumn{1}{c|}{} & \multicolumn{1}{l|}{Conv2D layers} & 3     &  \\
			\multicolumn{1}{c|}{} & \multicolumn{1}{l|}{Hidden size} & 192   &  \bigstrut[b]\\
			\hline
			
		\end{tabular}%
		\label{tab:model_configs}%
	\end{table*}%

	\subsection{Subjective Evaluation}
	\label{appendix:subjective_evaluation}
	For each tested dataset, we randomly select 10 texts from the test set and use the TTS systems to generate the audio samples. Each audio has been listened to by at least 20 native listeners, who are recruited on a crowdsourcing platform, Zhengshu Technology. We tell listeners to \textit{"focus on examing the naturalness of prosody (e.g., pitch, energy, and duration) and audio quality (noise, timbre, sound clarity, and high-frequency details)"}. For MOS, each tester is asked to evaluate the subjective naturalness of a sentence on a 1-5 Likert scale. For CMOS, listeners are asked to compare pairs of audio generated by systems A and B and indicate which of the two audio they prefer and choose one of the following scores: 0 indicating no difference, 1 indicating small difference, 2 indicating a large difference, and 3 indicating a very large difference. 
	
	\section{More Details in Analysis}
	\label{appendix:analysis}
	\subsection{Example Sentences}
	We list the 8 example sentences in Table \ref{tab:8_sentences}. These sentences are used as examples in Section \ref{sec:deeper_analysis}.
	\begin{table*}[t]
		\centering
		\resizebox{0.98\textwidth}{!}{%
			\begin{tabular}{l|l}
				\toprule
				s1 & \textit{... for the reputation of the stern judge stands not \textbf{higher} than that of the compassionate ...} \\
				s2 & \textit{As I went on , the precipices rose \textbf{higher} and seemed to overhang. The channel grew narrower ... }  \\
				s3 & \textit{Better, and better, and better! Her voice went \textbf{higher} with each better, till it got quite to a squeak at last.} \\
				s4 & \textit{... and the native graduates of our \textbf{higher} institutions have begun to show their strength ...}  \\
				s5 & \textit{Innocence is \textbf{higher} than virtue.}  \\
				s6 & \textit{Nothing seems more unfit to give a deeper meaning to life and a \textbf{higher} value.}  \\
				s7 & \textit{\textbf{Higher} up could be seen some chinamen, but whether they were fishing or washing we could not tell .}  \\
				s8 & \textit{May they become convalescents and overcomers, and create \textbf{higher} bodies for themselves !}  \\
				\bottomrule
			\end{tabular}
		}
		\caption{The text sentences used in the intuitive example, the selected word token \textit{"higher"} is bold.}
		\label{tab:8_sentences}
		\vspace{-5pt}
	\end{table*}
	
	\subsection{Self-similarity of Other Baselines}
\begin{figure}
		\centering
		\subfigure[A$^3$T]{
			\begin{minipage}[t]{0.47\linewidth}
				\centering
				\includegraphics[width=0.99\linewidth]{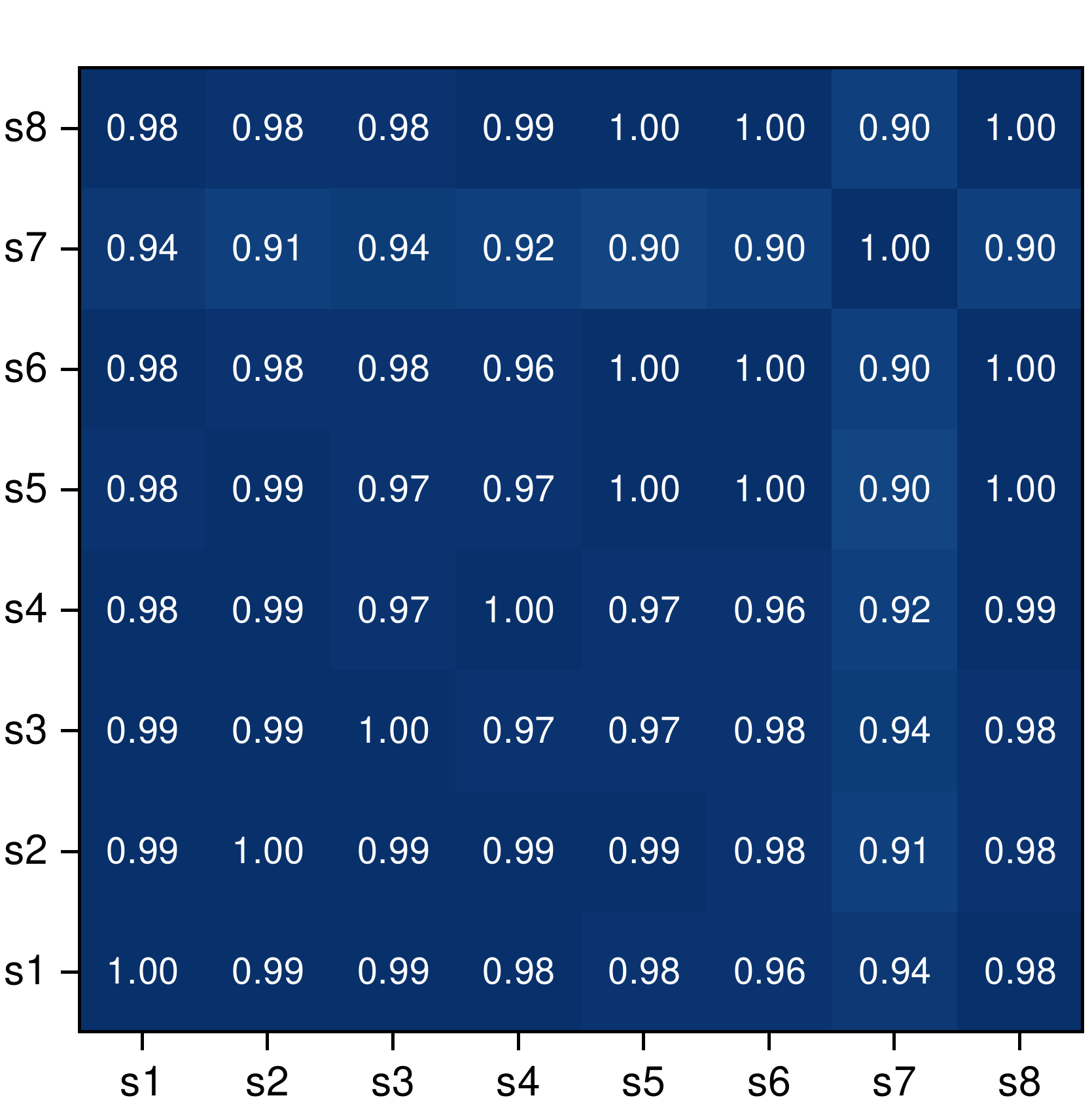}
			\end{minipage}
			\label{fig:heatmap_a3t}
		}%
		\subfigure[BERT]{
			\begin{minipage}[t]{0.51\linewidth}
				\centering
				\includegraphics[width=0.99\linewidth]{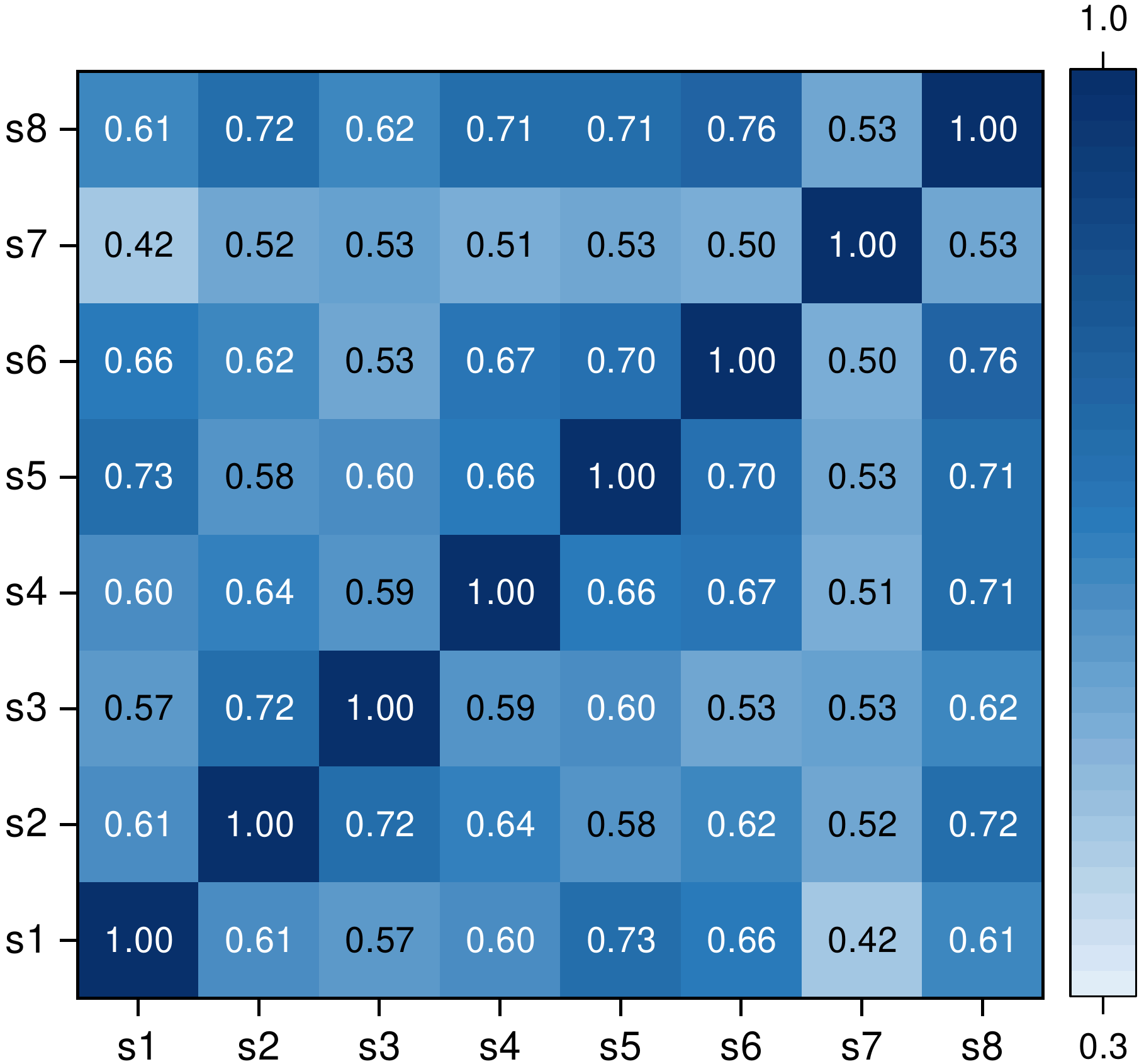}
			\end{minipage}
			\label{fig:heatmap_bert}
		}
		\caption{Self-similarity matrix visualization of A$^3$T and BERT.}
		\label{fig:heatmap2}
	\end{figure}

	The self-similarity visualization of A$^3$T and BERT can be found in Figure \ref{fig:heatmap2}. We discuss the results in Section \ref{sec:discuss_self_similarity}.
	
\end{document}